\documentclass[namedreferences,hyperref,optionalrh]{spr-sola}
\usepackage{graphicx}        
\usepackage{amsmath}
\usepackage{color}           

\newcommand{\aditya}{{Aditya-L1}}
\newcommand{\suit}{{SUIT}}
\newcommand{\suitc}{{Solar Ultraviolet Imaging Telescope (SUIT)}}
\newcommand{\vel}{{VELC}}
\newcommand{\velc}{{Visible Emission Line Coronagraph (VELC)}}
\newcommand{\sol}{{SoLEXS}}
\newcommand{\solc}{{Solar Low Energy X-ray Spectrometer (SoLEXS)}}
\newcommand{\hel}{{HEL1OS}}
\newcommand{\helc}{{High Energy L1 Orbiting Spectrometer (HEL1OS)}}
\newcommand{\viz}{{viz.}}
\newcommand{\aspex}{{Aditya Solar Particle EXperiment (ASPEX)}}
\newcommand{\papa}{{Plasma Analyser Package for Aditya (PAPA)}}
\newcommand{\magn} {{Magnetometers}}

\newcommand{\isro}{{ISRO}}

\newcommand{\arcsec}{{"}} 
\chardef\us=`\_
\graphicspath{{./}{figures/}}
\newcommand{\js}[1]{{\bf\color{blue} #1}}
\begin{document}
\begin{frontmatter}
\title{The Solar Ultraviolet Imaging Telescope on board Aditya-L1}
\author[addressref={iucaa,cessi},corref,email={durgesh@iucaa.in}]{\inits{Durgesh}\fnm{Durgesh}~\snm{Tripathi}\orcid{0000-0003-1689-6254}}
\author[addressref={iucaa,cessi},corref,email={anr@iucaa.in}]{\inits{A.R.}\fnm{A.N.}~\snm{Ramaprakash}\orcid{0000-0001-5707-4965}}
\author[addressref={manipal},,corref,email={sreejith.p@manipal.edu}]{\inits{S.P.}\fnm{Sreejith}~\snm{Padinhatteeri}\orcid{0000-0002-7276-4670}}
\author[addressref={iucaa,tezpur}]{\inits{J.S.}\fnm{Janmejoy}~\snm{Sarkar}\orcid{0000-0002-8560-318X}}
\author[addressref={iucaa}]{\inits{M.B.}\fnm{Mahesh}~\snm{Burse}}
\author[addressref={ursc}]{\inits{A.T.}\fnm{Anurag}~\snm{Tyagi}}
\author[addressref={iucaa}]{\inits{R.K.}\fnm{Ravi}~\snm{Kesharwani}\orcid{0009-0002-2528-5738}}
\author[addressref={iucaa}]{\inits{S.S.}\fnm{Sakya}~\snm{Sinha}}
\author[addressref={iucaa}]{\inits{B.J.}\fnm{Bhushan}~\snm{Joshi}}
\author[addressref={iucaa}]{\inits{R.D.}\fnm{Rushikesh}~\snm{Deogaonkar}\orcid{0009-0000-2781-9276}}
\author[addressref={iucaa}]{\inits{S.R.}\fnm{Soumya}~\snm{Roy}\orcid{0000-0003-2215-7810}}
\author[addressref={iucaa}]{\inits{V.N.}\fnm{VN}~\snm{Nived}}\orcid{0000-0001-6866-6608}
\author[addressref={iucaa}]{\inits{R.G.}\fnm{Rahul}~\snm{Gopalakrishnan}\orcid{0000-0002-1282-3480}}
\author[addressref={iucaa}]{\inits{A.K.}\fnm{Akshay}~\snm{Kulkarni}}
\author[addressref={iucaa}]{\inits{A.R.K.}\fnm{Aafaque}~\snm{Khan}\orcid{0000-0002-1244-0295}}
\author[addressref={iucaa,cessi}]{\inits{A.G.}\fnm{Avyarthana}~\snm{Ghosh}\orcid{0000-0002-7184-8004}}
\author[addressref={iucaa}]{\inits{C.R.}\fnm{Chaitanya}~\snm{Rajarshi}}
\author[addressref={iucaa}]{\inits{D.M.}\fnm{Deepa}~\snm{Modi}}
\author[addressref={ursc}]{\inits{G.K.}\fnm{Ghanshyam}~\snm{Kumar}} 
\author[addressref={ursc}]{\inits{R.Y.}\fnm{Reena}~\snm{Yadav}}
\author[addressref={iia}]{\inits{M.V.}\fnm{Manoj}~\snm{Varma}}
\author[addressref={uso}]{\inits{R.B.}\fnm{Raja}~\snm{Bayanna}}
\author[addressref={iucaa}]{\inits{P.C.}\fnm{Pravin}~\snm{Chordia}}
\author[addressref={cessi,iiserk}]{\inits{M.K.}\fnm{Mintu}~\snm{Karmakar}}
\author[addressref={iucaa}]{\inits{L.A.}\fnm{Linn}~\snm{Abraham}}
\author[addressref={manipal}]{\inits{H.A.}\fnm{H.N.}~\snm{Adithya}\orcid{0009-0002-1177-9948}}
\author[addressref={ursc}]{\inits{A.A.}\fnm{Abhijit}~\snm{Adoni}}
\author[addressref={tezpur}]{\fnm{Gazi A.} \snm{Ahmed}\orcid{0000-0002-0631-4831}}
\author[addressref={iia,aries}]{\inits{D.B.}\fnm{Dipankar}~\snm{Banerjee}}
\author[addressref={leos}]{\inits{B.R.}\fnm{Bhargava Ram}~\snm{B. S.}\orcid{0000-0001-7634-1790}}
\author[addressref={iucaa}]{\inits{R.B.}\fnm{Rani}~\snm{Bhandare}}
\author[addressref={iia}]{\inits{S.C.}\fnm{Subhamoy}~\snm{Chatterjee}}
\author[addressref={iucaa}]{\inits{K.C.}\fnm{Kalpesh}~\snm{Chillal}}
\author[addressref={ursc}]{\inits{A.D.}\fnm{Arjun}~\snm{Dey}}
\author[addressref={mps}]{\inits{A.G.}\fnm{Achim}~\snm{Gandorfer}}
\author[addressref={leos}]{\inits{G.G.}\fnm{Girish}~\snm{Gowda}}
\author[addressref={iisu}]{\inits{T.H.}\fnm{T. R.}~\snm{Haridas}}
\author[addressref={ursc}]{\inits{A.J.}\fnm{Anand}~\snm{Jain}}
\author[addressref={iucaa}]{\inits{M.J.}\fnm{Melvin}~\snm{James}}
\author[addressref={iisu}]{\inits{R.J.}\fnm{R. P. }~\snm{Jayakumar}}
\author[addressref={ursc}]{\inits{E.J.}\fnm{Evangeline Leeja}~\snm{Justin}}
\author[addressref={iia}]{\inits{N.K.}\fnm{Nagaraju}~\snm{K}}
\author[addressref={iucaa}]{\inits{D.K.}\fnm{Deepak}~\snm{Kathait}}
\author[addressref={iucaa}]{\inits{P.K.}\fnm{Pravin}~\snm{Khodade}}
\author[addressref={leos}]{\inits{M.K.}\fnm{Mandeep}~\snm{Kiran}}
\author[addressref={iucaa}]{\inits{A.K.}\fnm{Abhay}~\snm{Kohok}}
\author[addressref={mps}]{\inits{N.K.}\fnm{Natalie}~\snm{Krivova}}
\author[addressref={iisu}]{\inits{N.K.}\fnm{Nishank}~\snm{Kumar}}
\author[addressref={iucaa}]{\inits{N.M.}\fnm{Nidhi}~\snm{Mehandiratta}}
\author[addressref={iucaa}]{\inits{V.M.}\fnm{Vilas}~\snm{Mestry}}
\author[addressref={ursc}]{\inits{S.M.}\fnm{Srikanth}~\snm{Motamarri}}  
\author[addressref={ursc}]{\inits{S.M.}\fnm{Sajjade F.}~\snm{Mustafa}}
\author[addressref={cessi,iiserk}]{\inits{D.N.}\fnm{Dibyendu}~\snm{Nandy}}
\author[addressref={ursc}]{\inits{S.N.}\fnm{S.}~\snm{Narendra}}
\author[addressref={ursc}]{\inits{S.N.}\fnm{Sonal}~\snm{Navle}}
\author[addressref={ursc}]{\inits{N.P.}\fnm{Nashiket}~\snm{Parate}}
\author[addressref={ursc}]{\inits{A.P.}\fnm{Anju}~\snm{M Pillai}}
\author[addressref={iucaa}]{\inits{S.P.}\fnm{Sujit}~\snm{Punnadi}}
\author[addressref={ursc}]{\inits{R.A.}\fnm{A.}~\snm{Rajendra}}
\author[addressref={ursc}]{\inits{A.R.}\fnm{A.}~\snm{Ravi}}
\author[addressref={leos}]{\inits{B.S.}\fnm{Bijoy}~\snm{Raha}}
\author[addressref={ursc,cessi}]{\inits{K.S.}\fnm{K.}~\snm{Sankarasubramanian}}
\author[addressref={ursc}]{\inits{G.S.}\fnm{Ghulam}~\snm{Sarvar}}
\author[addressref={ursc}]{\inits{N.S.}\fnm{Nigar}~\snm{Shaji}}
\author[addressref={ursc}]{\inits{N.S.}\fnm{Nidhi}~\snm{Sharma}}
\author[addressref={iucaa}]{\inits{A.S.}\fnm{Aditya}~\snm{Singh}}
\author[addressref={ursc}]{\inits{S.S.}\fnm{Shivam}~\snm{Singh}}
\author[addressref={mps}]{\inits{S.S.}\fnm{Sami K.}~\snm{Solanki}}
\author[addressref={ursc}]{\inits{V.S.}\fnm{Vivek}~\snm{Subramanian}}
\author[addressref={ursc}]{\inits{R..}\fnm{Rethika}~\snm{T}}
\author[addressref={ursc}]{\inits{S.T.}\fnm{Srikanth}~\snm{T}}
\author[addressref={ursc}]{\inits{S..}\fnm{Satyannarayana }~\snm{Thatimattala}}
\author[addressref={ursc}]{\inits{H.T.}\fnm{Hari Krishna}~\snm{Tota}}
\author[addressref={ursc}]{\inits{V.T.}\fnm{Vishnu}~\snm{TS}\orcid{0000-0003-4892-3739}}
\author[addressref={iucaa}]{\inits{A.U.}\fnm{Amrita}~\snm{Unnikrishnan}\orcid{0000-0002-8152-9023}}
\author[addressref={ursc}]{\inits{K.V.}\fnm{Kaushal}~\snm{Vadodariya}}
\author[addressref={ursc}]{\inits{D.V.}\fnm{D. R.}~\snm{Veeresha}}
\author[addressref={leos}]{\inits{R.V.}\fnm{R}~\snm{Venkateswaran}}
\address[id=iucaa]{Inter-University Centre for Astronomy and Astrophysics, Post Bag 4, Ganeshkhind, Pune - 411007, Maharashtra, India}
\address[id=cessi]{Center of Excellence in Space Sciences India, Indian Institute of Science Education and Research Kolkata, Mohanpur 741246, West Bengal, India}
\address[id=manipal]{Manipal Centre for Natural Sciences, Manipal Academy of Higher Education, Karnataka, Manipal- 576104, India}
\address[id=tezpur]{Department of Physics, Tezpur University, Napaam, Tezpur-784028, Assam, India}
\address[id=ursc]{U R Rao Satellite Centre,  Old Airport Road Vimanapura Post, Bengaluru - 560017, Karnataka, India}
\address[id=iia]{Indian Institute of Astrophysics, Koramangala, Bengaluru - 560034, Karnataka, India}
\address[id=uso]{Udaipur Solar Observatory (USO), Udaipur, Rajasthan, India}
\address[id=iiserk]{Department of Physical Sciences, Indian Institute of Science Education and Research Kolkata, Mohanpur 741246, West Bengal, India}
\address[id=aries]{Aryabhatta Research Institute of Observational Sciences (ARIES), Manora Peak, Nainital - 263001 Uttarakhand, India}
\address[id=leos]{Laboratory for Electro-Optics Systems (LEOS), ISRO, First Cross, First Phase, Peenya, Bengaluru- 560058, Karnataka}
\address[id=mps]{Max Planck Institute for Solar System Research, Justus-von-Liebig-Weg 3, 37077 G\"ottingen, Germany}
\address[id=iisu]{ISRO Inertial Systems Unit, Vattiyoorkavu Complex, Nettayam, Thiruvananthapuram- 695022, Kerala}
\runningauthor{Tripathi et al.}
\runningtitle{SUIT on board Aditya-L1}
\begin{abstract}
The Solar Ultraviolet Imaging Telescope (SUIT) is an instrument on the Aditya-L1 mission of the Indian Space Research Organization (ISRO) launched on September 02, 2023. SUIT continuously provides, near-simultaneous full-disk and region-of-interest images of the Sun, slicing through the photosphere and chromosphere and covering a field of view up to 1.5 solar radii. For this purpose, SUIT uses 11 filters tuned at different wavelengths in the 200{--}400~nm range, including the Mg~{\sc ii} h~and~k and Ca~{\sc ii}~H spectral lines. The observations made by  SUIT help us understand the magnetic coupling of the lower and middle solar atmosphere. In addition, for the first time, it allows the measurements of spatially resolved solar broad-band radiation in the near and mid ultraviolet, which will help constrain the variability of the solar ultraviolet irradiance in a wavelength range that is central for the chemistry of the Earth's atmosphere. This paper discusses the details of the instrument and data products. 
\end{abstract}

\end{frontmatter}
\section{Introduction}\label{S-Introduction} 
\begin{figure*}[!ht]
\centering
\includegraphics[width=0.7\textwidth]{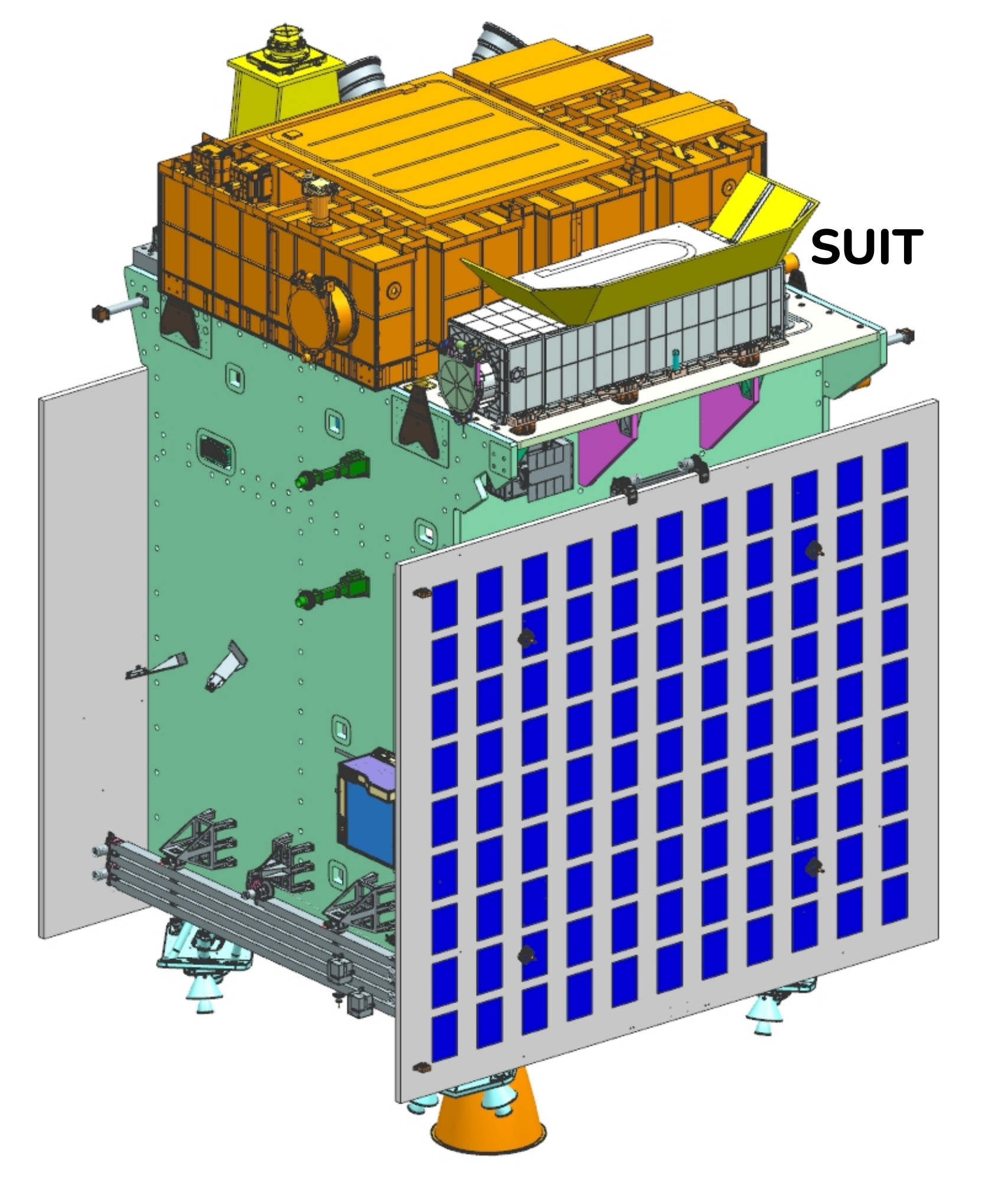}
\caption{A CAD model of the {\aditya} spacecraft. \suit is labeled. \label{fig:sc} }
\end{figure*}

Aditya-L1 is the first space-based solar observatory mission of the {\isro}. The mission's primary goal is to study the coupling and dynamics of the magnetized solar atmosphere using multi-wavelength observations and the properties of the solar wind at the first Lagrange point using in-situ measurements. {\aditya} carries seven instruments to achieve its goals, comprising remote sensing and in-situ instruments \citep[][]{aditya_mission}. The remote sensing instruments cover a wavelength range from Hard X-ray (HXR) to infra-red wavelength using four different instruments {\viz} the {\velc}, the {\suitc}, the {\helc} and {\solc}. The in-situ instruments cover the energy range from 0.01{--}20~keV using two instruments {\viz} {\aspex} and {\papa}. Finally, the two identical {\magn} mounted on a boom continuously monitors the interplanetary magnetic field. Figure~\ref{fig:sc} displays a CAD model of {\aditya} spacecraft mounted with all the seven instruments.

{\suit}, mounted on the top deck of the satellite (as labelled in Figure~\ref{fig:sc}), is one of the four remote sensing instruments designed to help {\aditya} meet its goals.  It is a combined medium and narrow-band filter imager of the Sun in the Near Ultraviolet (NUV) wavelength band {\viz} 200{--}400~nm. On the one hand, the radiation in this wavelength band comes from the photosphere and chromosphere, which are critical for addressing the issues related to the magnetic coupling of the entire solar atmosphere. On the other hand, it provides a unique opportunity to measure and monitor the contributions of various spatial structures to the variations of Solar Spectral Irradiance (SSI) in the NUV, which is crucial for the chemistry of Ozone and Oxygen within the stratosphere of the earth and, thereby, the Sun-climate relationship.

The observations recorded by SUIT will help us address a broad range of science questions \citep[][]{csp_suit}, focusing on the following questions:

\begin{itemize}
	\item Dynamic coupling of the lower solar atmosphere: what is the nature of MHD waves in different features such as quiet Sun, active regions and sunspots in photosphere and chromosphere and what role do these waves play in transferring the energy from the photosphere to the chromosphere.
	\item Sun-climate relationship: Why and how strongly does the Solar Spectral Irradiance in the NUV, which is relevant for Earth’s climate, vary?
	\item Solar flare dynamics and their energy distribution: at what wavelength do flares radiate most of their energy and what fraction of a flare's energy is contained within the NUV range? What is the spectral energy distribution in flares?
	\item Physics of eruptions at various spatio-temporal scales: what are chromospheric signatures and counterparts of eruptive phenomena such as jets, macro-spicules etc. occurring at diverse spatio-temporal scales?
\end{itemize}

It is essential to highlight that so far there have been only limited imaging facilities within this wavelength band, {\viz} the Ultraviolet Spectrometer and Polarimeter \citep[UVSP;][]{uvsp} on the Solar Maximum Mission \citep[SMM;][]{SMM}, the Interface Region Imaging Spectrometer \citep[IRIS;][]{iris} and the two flights of the Sunrise mission \citep[][]{sunrise_sami, sunrise_barthol, sami_sunrise}. The UVSP and IRIS have a very limited spectral and spatial coverage. Though for a very small duration, the SuFI instrument \citep{SuFI} on Sunrise I and II, however, did cover the  wavelength range from 200{--}400~nm but had a very small field of view (FOV) with an extremely high spatial resolution ($\approx$~0.02 arcsec/pixel). Additionally, in the Sunrise III flight the Sunrise Ultraviolet Spectrograph and Imager \citep[SUSI,][]{susi} instrument also covered 309~nm, albeit at extremely high resolution and a small FOV.

We note that, albeit with limited coverage, the observations recorded with the above-mentioned facilities have demonstrated the capabilities of this wavelength window in terms of enhancement of our understanding of the solar atmosphere. Moreover, it is paramount to measure the spatially resolved contributions to solar spectral irradiance within this wavelength band to fully gauge the effects of solar irradiance variations on Earth's atmosphere \citep{KriSF_2003, ErmMD_2013, SolKS_2013}. To this end, SUIT observes the photosphere and chromosphere of the Sun and provides full-disk as well as region-of-interest images of the Sun in the wavelength range of 200{--}400~nm for the first time. The SUIT instrument on board {\aditya} mission opens up an unprecedented observing window for the Sun at NUV wavelengths, without attenuation due to Earth's atmosphere. 

Combining the observations taken from SUIT with those from the {\vel}, {\sol} and {\hel}, we will have complete coverage of the solar atmosphere from the photosphere to the corona and a broad temperature coverage from a single platform provided by {\aditya}. Moreover, combining the observations from SUIT with those taken from instruments on \citep[Hinode][]{hinode} the Atmospheric Imaging Assembly \citep[AIA;][]{aia} and the Helioseismic and Magnetic Imager \citep[HMI;][]{hmi}, both on board the Solar Dynamics Observatory \citep[SDO;][]{sdo} and the Extreme Ultraviolet Imager \citep[EUI;][]{eui} and the Polarimetric and Helioseismic Imager \citep[PHI;][]{phi} on board Solar Orbiter \citep[][]{sorbiter} will provide unprecedented opportunities to study the Sun in great detail.

This paper presents the technical details of the {\suit} and its various components, including their performance. The rest of the paper is arranged as follows. In Section\ref{I-overview}, we present the overview of the instrument. The optical design and imaging performance are presented in Section~\ref{S-Optical_des}. We discuss the development and selection of different components in Section~\ref{S-components}. The four mechanisms in \suit are discussed in Section~\ref{S-mechanisms}. We discuss the SUIT electronics package in Section~\ref{S-PEPackage}, structural design and analysis in Section~\ref{S-Structural} and thermal control system in Section~\ref{S-Thermal}. Finally, we describe the data analysis pipeline, data management and distribution in Section~\ref{S-Data_Management}, followed by a summary in Section~\ref{S-conclusions}.

\begin{figure*}
\centering
\includegraphics[width=0.50\textwidth]{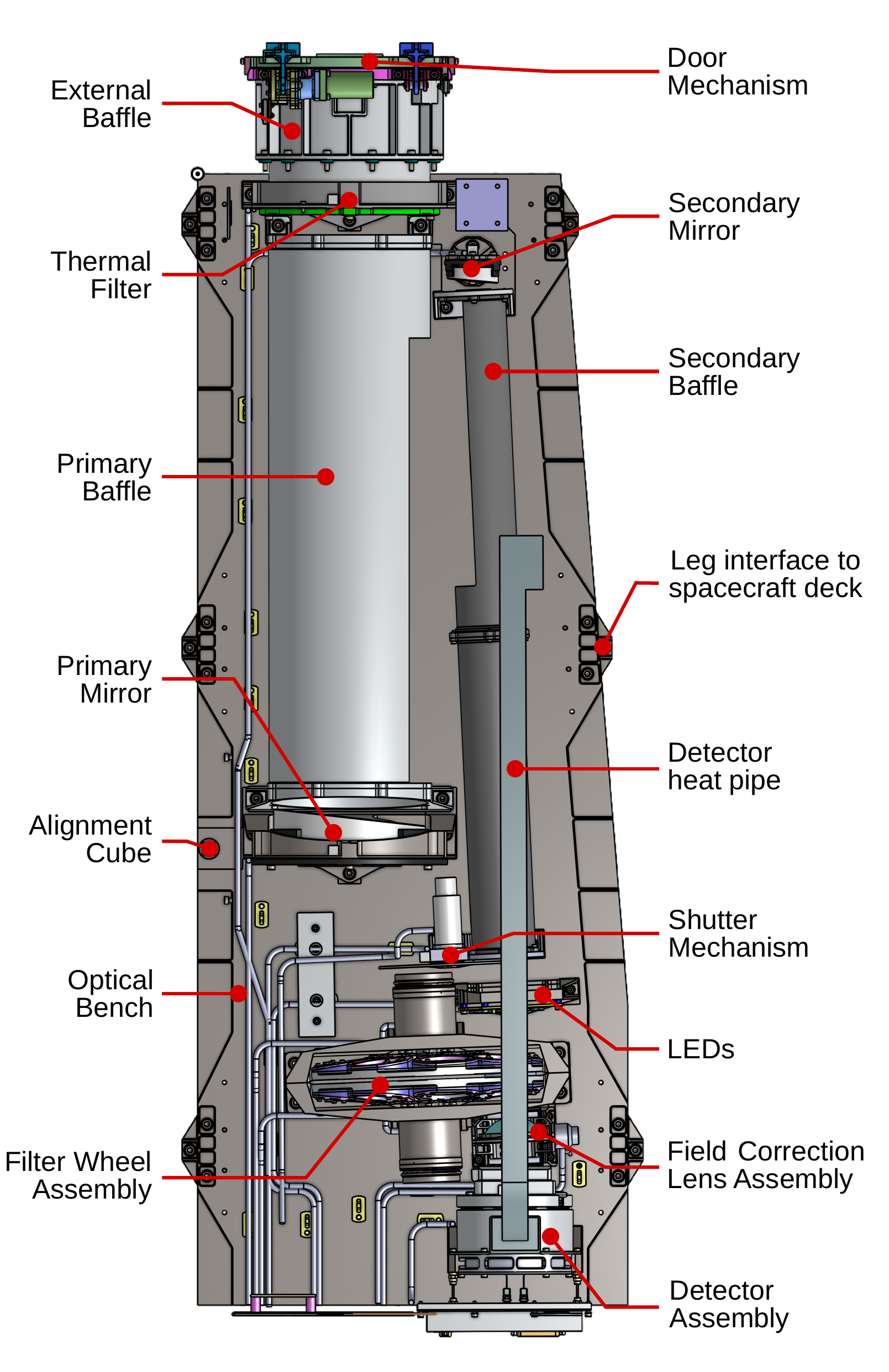}
\includegraphics[width=0.4\textwidth]{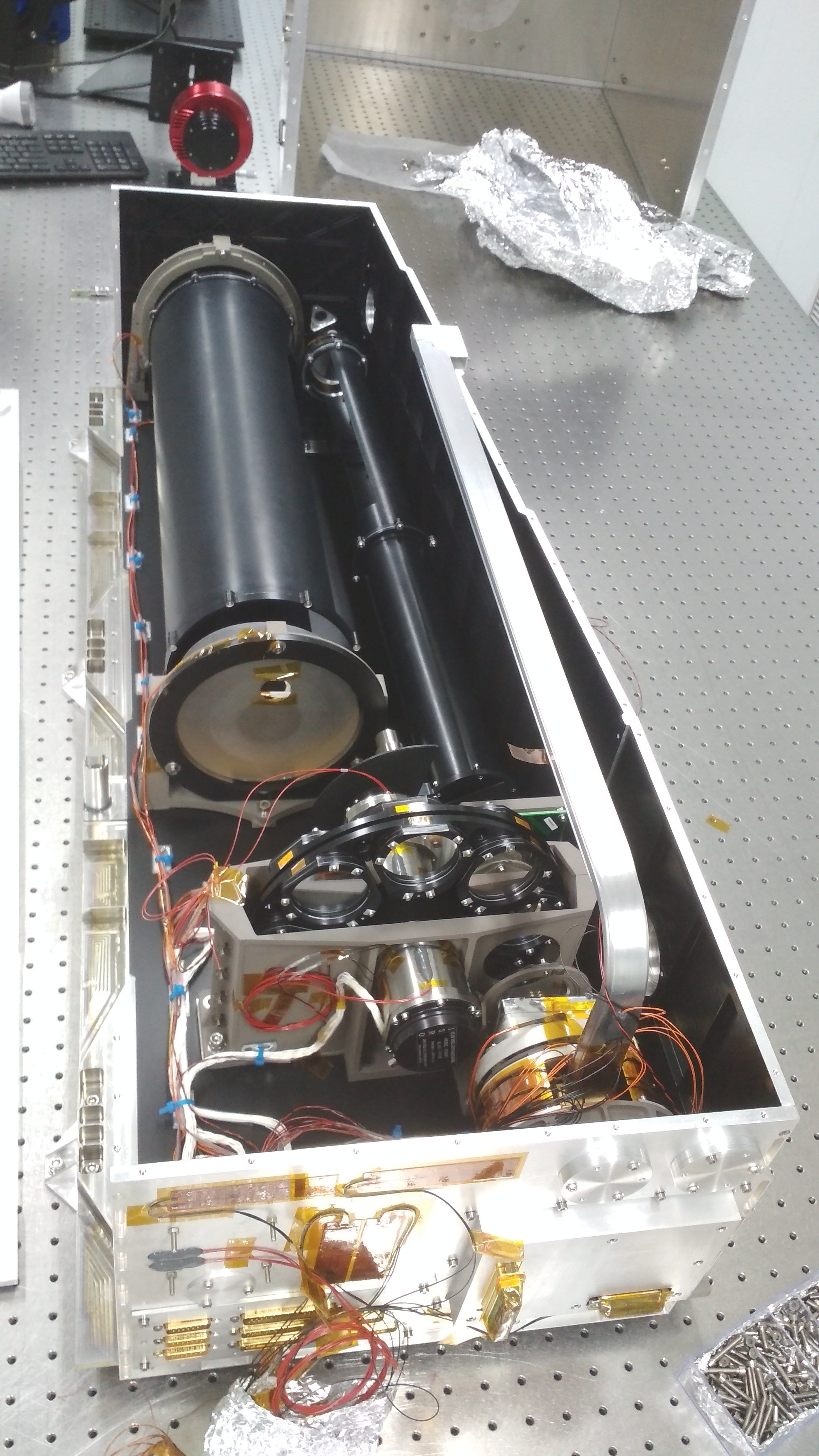}
\caption{Left: CAD model of SUIT assembly with the top cover and radiator assembly removed, showing the internal configuration of the telescope. Various components of the telescopes are labeled. Right: \suit without the top cover panel.}\label{fig:layout}
\end{figure*}

\section{Instrument Overview}\label{I-overview}
The SUIT instrument is a two-mirror f/24.8 off-axis Ritchey–Chrétien telescope with an effective focal length of 3500~mm. It is designed and optimised to take images of the Sun with a resolution of $\approx$~1.4{\arcsec}, and a field of view (FOV) of 1.5~R$_{\odot}$ on a UV-enhanced back illuminated 4096~$\times$~4096 CCD with 12 $\mu$m size pixels. 

\suit is located on the top deck of the {\aditya} spacecraft as depicted in Figure~\ref{fig:sc}. It has three primary sub-units, viz. the telescope assembly, the payload electronics (PE) package, and the filter wheel drive electronics (FWE) package. The PE and FWE packages are located on the intermediate deck of the spacecraft. The PE package consists of the drive electronics for the CCD, the shutter, and the focusing mechanisms. The telescope assembly and the FWE are connected to the PE via SpaceWire harness.

\begin{table*}[!ht]
\caption{Filter abbreviations (Mnemonics), central wavelength, and bandpass of the 11 science filters for SUIT.} \label{sc_comb_fil}
\begin{tabular}{lcccr}
\hline
Filter ID 	& Central 		    & FWHM 		& Combination		 & Remarks\\
		& Wavelength	    &			& Filter			&         \\
		& (nm)			    & (nm)		&					&          \\
\hline
NB1 		& 214.0             & 11  	    & BB1    			&Continuum\\
NB2 		& 276.7 		    & 0.4  		& BPF1   			&Continuum\\
NB3 		& 279.6 			& 0.4  		& BPF1   			&Mg~{\sc ii}~k\\
NB4 		& 280.3 			& 0.4		& BPF1   			&Mg~{\sc ii}~h\\
NB5 		& 283.2 			& 0.4		& BPF1   			&Continuum\\
NB6 		& 300.0   			& 1.0		& BPF2   			&Continuum\\
NB7 		& 388.0   			& 1.0		& BPF2   			&CN Band\\ 
NB8 		& 396.85 			& 0.1 		& NB8  				&Ca~{\sc ii}~h\\
BB1 		& 200{--}242 		& 42.0		& BB1   			&Herzberg Continuum\\
BB2 		& 242{--}300 		& 58.0		& BPF3   			&Hartley Band\\
BB3 		& 300{--}360 		& 40.0		& BPF3   			&Huggins Band\\
\hline
\end{tabular}
\end{table*}

\begin{table*}[!ht]
\caption{SUIT Instrument Characteristics}\label{tab:instrument}      
\begin{tabular}{l|r}
\hline
\textbf{Parameter} 							& \textbf{Value}   \\
\hline
\hline
\textbf{Telescope}                          &\\					
Design                                      & f/24.8 two mirror off-axis design \\
Entrance Aperture 							& 146 mm \\
Primary mirror  diameter                    & 140.8 mm\\
Focal Length 								& 3500 mm\\
&\\
\textbf{Detector}                           &\\
Type                                        & 4096 $\times$ 4096, back-thinned,\\ 
                                            & back-illuminated, UV-enhanced CCD\\
&\\
Full-well capacity 				            & 195,000 e$^{-}$ \\
Plate Scale			 						& 0.7 arcsecond/pixel @ 12 $\mu$m pixel\\
Field of View (FOV) 						&  2860$\times$ 2860 arcsecond$^{2}$ \\ 
&\\
\textbf{Cadence}                            & \\
Full Frame                                  & 20~s\\
Region of Interest (RoI)                    & 4~s\\
&\\
\textbf{Filters}                            & \\
Entrance aperture                           & blocking (out band)\\
Science filters                             & 8 Narrow-band, 3 Broad-band \\
&\\
Exposure times 								& 0.1 to 1.4 s  \\
Mass  										& 42 Kg\\
Power  										& 35 W\\ 
Daily Data Volume 							& 42 GB,  uncompressed\\ 
Compression      							& Factor of 2,  Lossless\\ 
\hline
\end{tabular}
\end{table*}

Figure~\ref{fig:layout} displays the top view of the telescope assembly without the top cover and radiator assembly. The entrance aperture of {\suit} consists of a multi-operational entrance door and a Thermal Filter (TF) assembly. The TF is designed such that it blocks most of the visible ($\approx$~99.75\%) and infrared radiation ($\approx$~99.5\%) and transmits only a fraction of the NUV radiation ($\approx$~0.3\%). In open-door conditions, the solar radiation enters the instrument optical cavity through the TF. The door is closed during the ground operation, launch, and transit. 

The light transmitted through the TF bounces off the primary and the secondary mirrors and passes through a combination of two filters. {\suit} has 11 science filters {\js (see Table~\ref{sc_comb_fil})} and five combination filters mounted on two filter wheels, each having 8 slots. The science filters are designed to meet specific scientific objectives. The combination filters are designed to limit the photon flux within the dynamic range of the CCD, combined with specific science filters. A desired combination of a science filter and a combination filter is achieved by moving the two independent filter wheels. The light that passes through the filters is focused on the detector by a field corrector lens. The lens is mounted on a linear piezo-driven stage to allow for possible focus adjustments during the instrument's in-orbit calibration. 

The PE box executes the operational modes, including the on board intelligence for flare detection, localization and tracking, and provides interfaces with spacecraft subsystems and other instruments. The filter wheel electronics (FWE) package operates the two filter wheel drive mechanisms based on the sequencing supplied by the PE box. We provide the detailed instrument specifications in Table~\ref{tab:instrument}.

\section{Optical Design and Imaging Performance}\label{S-Optical_des} 
\begin{figure*}
\centering
\includegraphics[trim=5cm 0cm 5cm 0cm,width=0.5\textwidth]{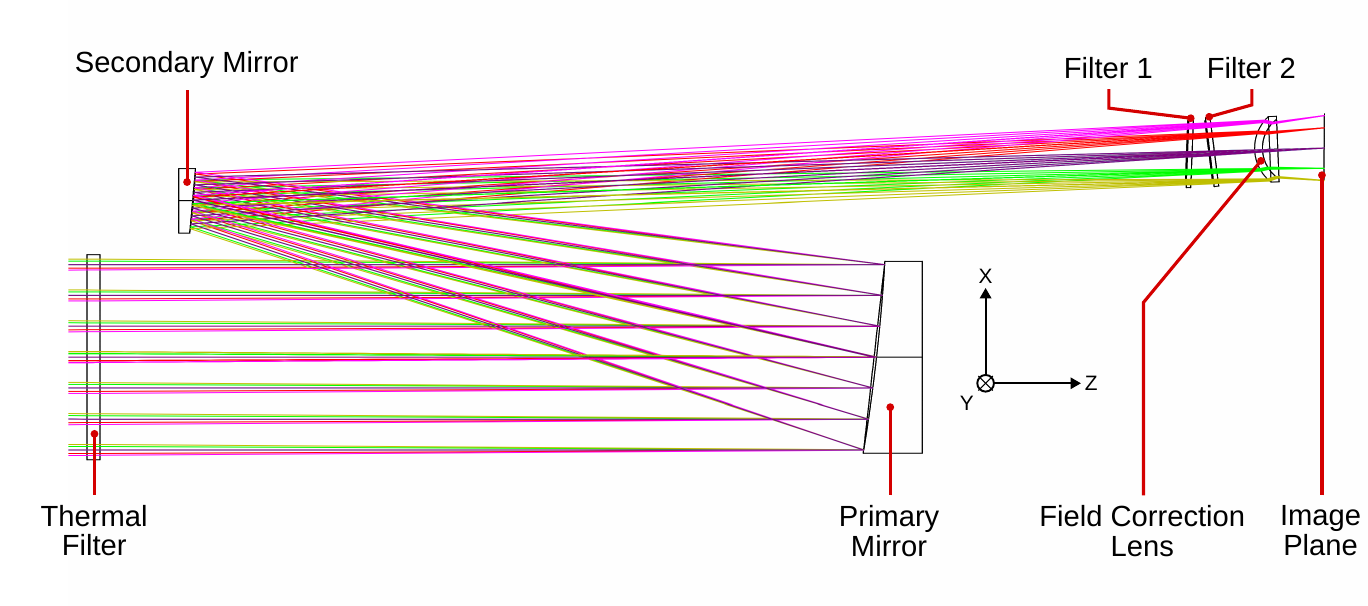}
\caption{2D layout of SUIT optical design. The tilted filter configuration has been derived from ghost image analysis. The reference axes is shown for defining tilts. Light rays of different colors show incident light of various field angles.} \label{fig:opt_des}
\end{figure*}

SUIT is an off-axis Ritchey Chr\'etien telescope as shown in Figure~\ref{fig:opt_des}. This design was chosen to accommodate the instrument within the available volume on the spacecraft deck. The off-axis design of the system also minimizes the scattering of light due to secondary optical elements and their mechanical support structure. The optical surfaces of the two mirrors, concave for primary and convex for secondary, have hyperbolic profiles optimized to balance the optical aberrations. A field corrector lens in front of the detector is designed to maintain the uniformity of image quality across the FOV. The field corrector lens is mounted on a focusing mechanism that can be used to compensate for image degradation due to first-order thermal effects in orbit.  The optical design parameters are given in Table~\ref{tab:opt}. The design has been optimized to provide 1.4~arcsec angular resolution across the 2860$\times$ 2860 sq. arcsecond FOV, with an effective focal length of 3500 mm at 280~nm.

\begin{table}[h!]
\caption{Optical Design parameters. Abbreviations are as follows {--} TF: Thermal Filter, PM: Primary Mirror, SM: Secondary Mirror}.\label{tab:opt}
\begin{tabular}{llr}
\hline
Component(s)            & Parameter             & Value\\
\hline
TF     & Clear Aperture        & 151 mm \\ 
                        & Thickness             & 10 mm \\
                        &\\
TF and PM   & Separation (Z)        & 600.80 mm \\ 
&\\
PM          & Clear Aperture        & 141 mm \\ 
                        & Radius of Curvature   &  -1404.67 mm \\
                        & Conic Constant        &  -1.0255 \\
                        & Off-Axis Distance     &  159 mm\\
&\\
PM and SM               &  Separation (Z)       &   527.806~mm\\ 
&\\
SM        & Clear Aperture        & 44 mm \\ 
                        & Radius of Curvature   &  -438.252 \\
                        & Conic Constant        &  -2.4867 \\
                        & Off-Axis Distance     &  40 mm\\
&\\
Filters                 & Clear Aperture        &  49 mm \\ 
                        & Thickness             &  3.5 mm \\
                        & Filter1 Tilt          &  Table~\ref{tab:filter_tilt} \\ 
                        & Filter2 Tilt          &  Table~\ref{tab:filter_tilt}\\
&\\
Lens                    & Clear Aperture        & 49 mm \\ 
                        & Thickness             & 6 mm \\
                        & Radius of Curvature (Front) &  32.971 mm \\
                        & Radius of Curvature (Back) &  28.0369 mm\\
                        & Tilt &  2.3432 degree \\
\hline
\end{tabular}
\end{table}

The thermal filter at the entrance aperture is optically flat and has a custom coating designed to suppress light outside the 200{--}400 nm wavelength region. The selection of science-specific band-pass is done by combining two filters in the beam path, mounted on two filter wheels. Each filter wheel has eight slots and the eleven science filters and five complementary band-pass filters are distributed on the wheels to minimize the time required to cycle through the filter combinations. The coatings for the science filters are designed to select the band-pass of scientific interest while the coatings for the complementary pass filters are designed to suppress the out-of-band light for the given science filter. The filter combinations are designed to optimize the photometric performance while maintaining the flux at the detector within the dynamic range.

The close proximity of the field corrector lens and the filters to the CCD affects the image quality by creating spurious patterns called "ghost" due to inter-reflections from surfaces. The major contribution of the ghost is due to the surface reflections from the CCD, which are negatively correlated with quantum efficiency. Along with the CCD coating thickness, the filter tilts (Table \ref{tab:filter_tilt}) are optimized to generate the least ghost flux within the FOV. 

\begin{figure*}[h]
\centering
\includegraphics[width=0.99\textwidth]{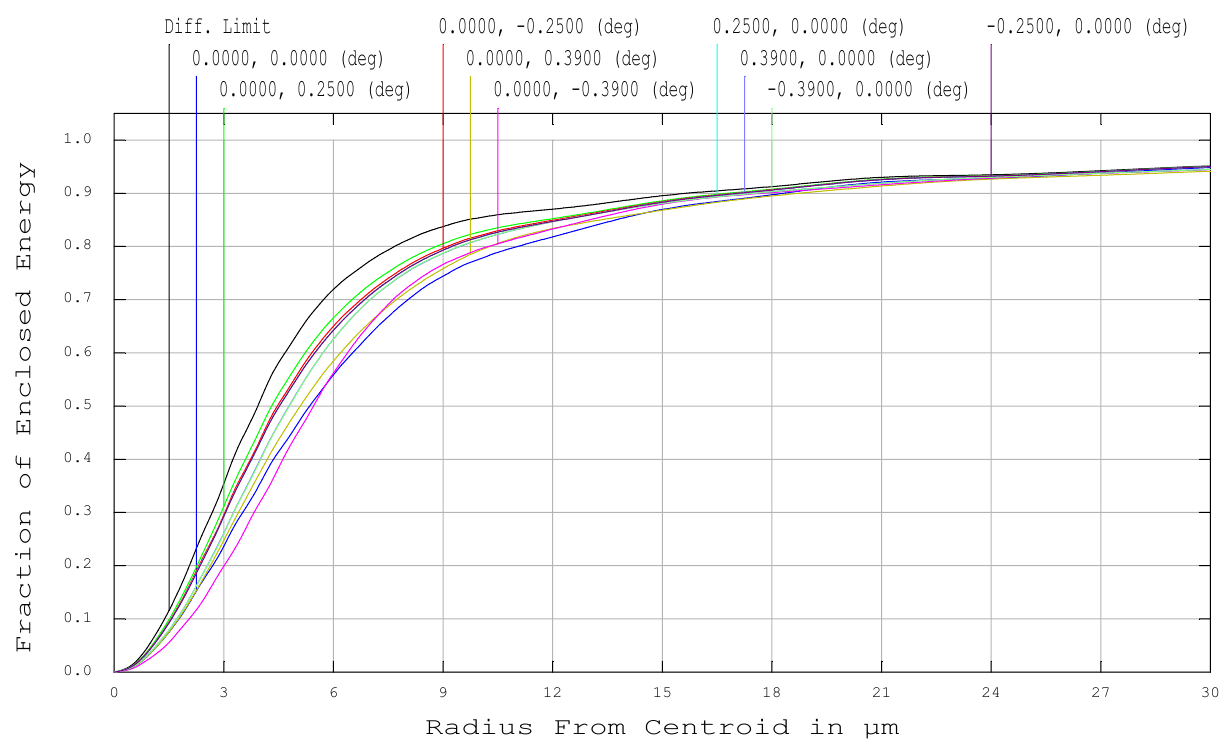}
\caption{Cumulative probability plot for SUIT 80\% diffraction scaled encircled energy after applying the tolerance in CODE-V.} \label{cumul} 
\end{figure*}

The tolerance analysis used Code V Optical Design Software to put bounds on the fabrication and alignment errors. For fabrication tolerances, the radius of curvatures, mirror off-axis distance, conic constant, filter wedge, and surface irregularity are considered. The element tilts and de-centers are considered for alignment tolerances. The wave-front differential method in Code V is used to estimate the effect of simultaneous errors, and the change in 80\% encircled energy diameter (Figure~\ref{cumul}) at different field points is used as a performance metric. The tolerance analysis is constrained to limit the combined effect of the errors such that the encircled energy diameter remains within 2 pixels, i.e., $\approx 24$ microns. The primary-secondary mirror separation, secondary mirror de-center, tilt, and lens-to-image plane distance are used to optimize for the optical performance within the acceptable limit.
\begin{table}[h!]
\caption{Science Filter Tilts.}\label{tab:filter_tilt}
\begin{tabular}{cccc}
\hline
Filter on FW1   & Tilt (deg)  &Filter on FW2   & Tilt (deg)\\
\hline
NB04    & 5 & BP02  & 6 \\
NB03    & 5 & NB06  & 4 \\
NB02    & 6 & NB07  & 4 \\
NB05    & 5 & NB08  & 0 \\
BP03    & 5 & BB01  & 4 \\
NB08    & 0 & NB01  & 6 \\
BB01    & 4 & BB02  & 6 \\
BP04    & 4 & BB03  & 4 \\
\hline
\end{tabular}
\end{table}

\begin{figure}
\centering
\includegraphics[width=0.7\textwidth]{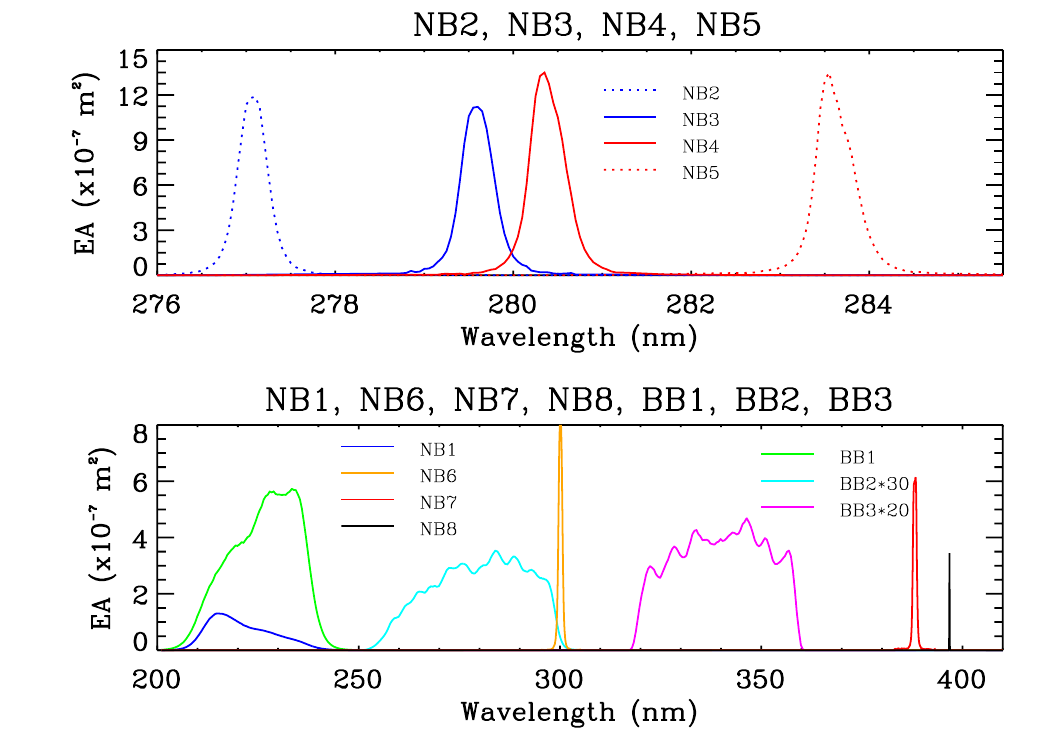}
\caption{Effective area as a function of wavelength of different filter combinations.} \label{fig:effective_area} 
\end{figure}

Figure~\ref{fig:effective_area} shows the effective area as a function of wavelength for the different filter combinations. The effective area includes the response of all the optical components along the ray path. So, the effective area is given by,

\begin{align*}
EA(\lambda) = & A~[TF(\lambda)~PMR(\lambda)~SMR(\lambda)~SF(\lambda)~CF(\lambda)~L(\lambda)~QE(\lambda)]\\
\noindent where, \\
& \mathrm{A = Area~of~the~entrance~aperture~=~0.01561~m^{2},} \\
& \mathrm{TF(\lambda) = Thermal~filter~transmission~profile,} \\
& \mathrm{PMR(\lambda) = Primary~mirror~reflectivity,} \\
& \mathrm{SMR(\lambda) = Secondary~mirror~reflectivity,} \\
& \mathrm{SF(\lambda) = Science~filter~transmission~profile,} \\
& \mathrm{CF(\lambda) = Complementary~filter~transmission~profile}, \\
& \mathrm{L(\lambda) = Focusing~mechanism transmission~profile}, \\
& \mathrm{QE(\lambda) = Quantum~efficiency} \\
\end{align*}

\begin{figure}[H]
    \centering
    \includegraphics[trim={0.8cm 0cm 2cm 1.5cm}, clip, width=0.7\textwidth]{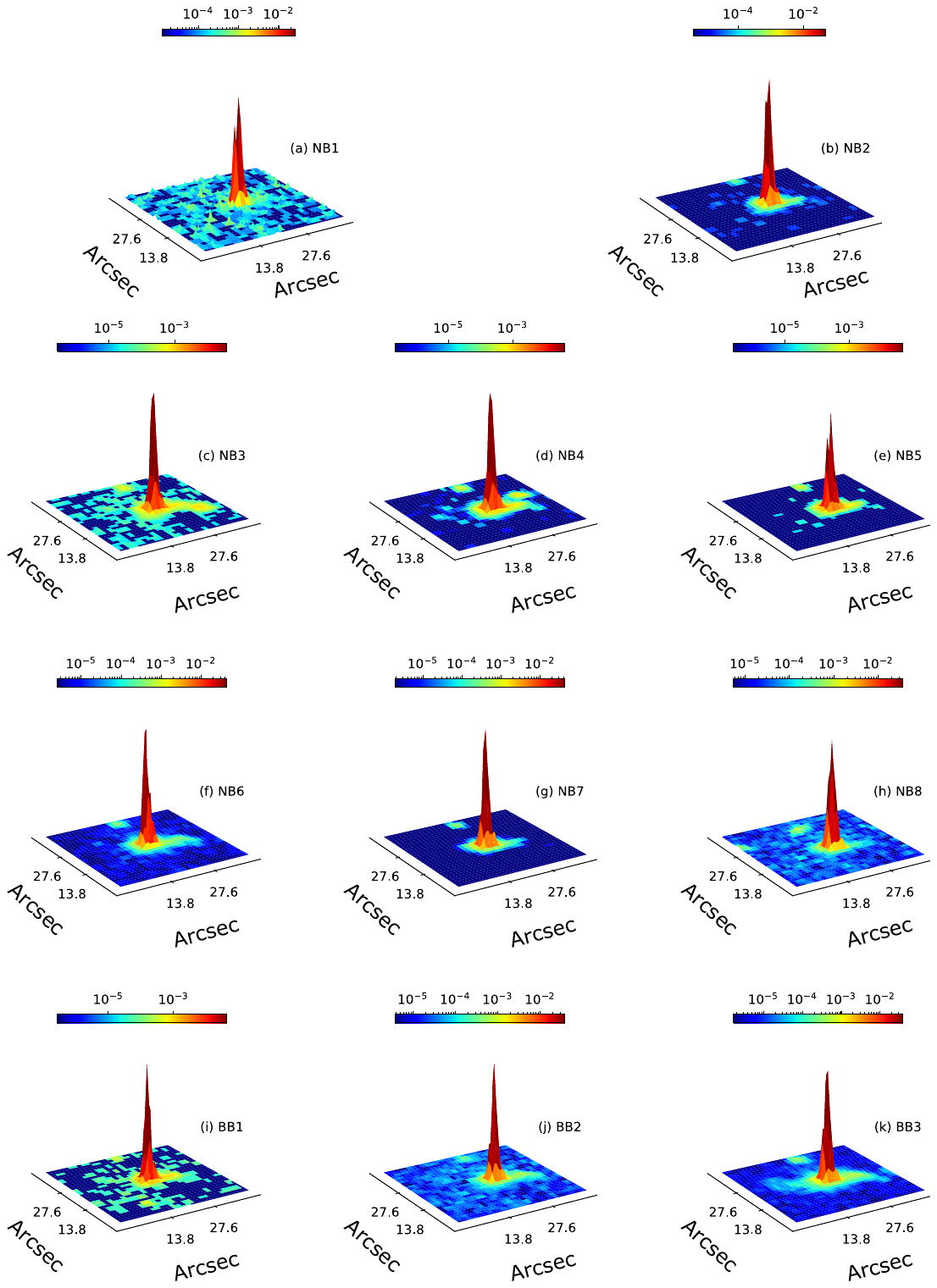}
    \caption{Measured SUIT PSF in detector centre for the various science filter combinations. The colorbar is normalized over the plotted region to add up to unity. The shape of the peak represents the shape of the PSF, where the height corresponds to the normalized intensity which is also depicted by the colormap.}
    \label{fig:psf}
\end{figure}

We perform an end-to-end optical test of the assembled instrument by feeding it with collimated NUV light. We record the PSF of \suit\ for the eleven science filter combinations at the center of the CCD in Figure~\ref{fig:psf}. The PSF patches are plotted centered around the central PSF peak in a $\approx~(40"~\times~40")$ box. The color map of the measured PSF is normalized over the plotted region to add up to unity. The color bar accompanying every panel shows the normalized colormap of the measured PSF. The shape of the PSF peak gives a clear picture of the PSF geometry, where the height corresponds to the normalized intensity, also depicted by the colormap.

\section{Development and Selection of Components}\label{S-components} 
\subsection{Thermal Filter (TF)} \label{S-Thermalfilter}
\begin{figure}
\centering
\includegraphics[width=0.8\textwidth]{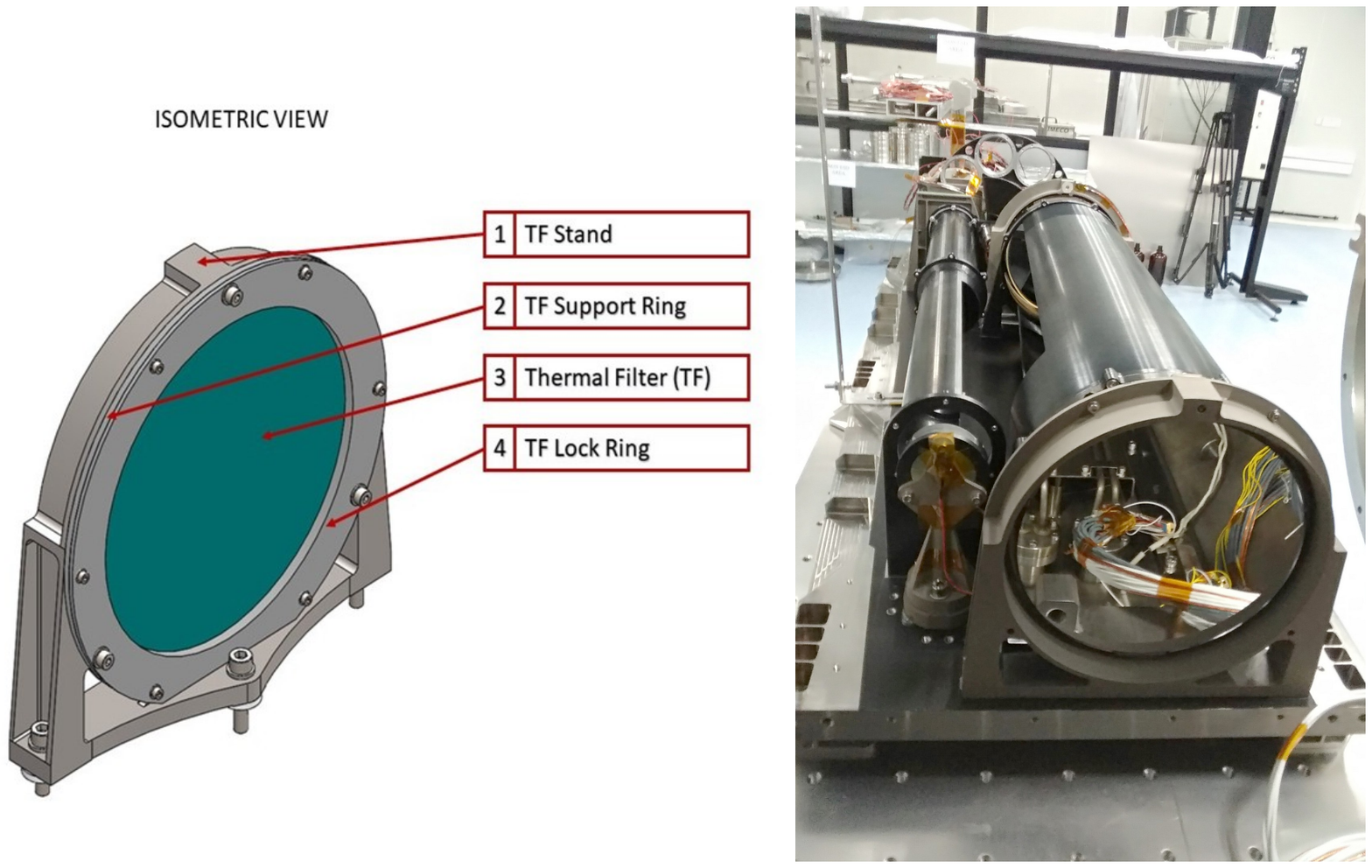}
\caption{Isometric view (left) and actual photo (right) of the thermal filter assembly mounted on the SUIT optical bench.}\label{fig:TF_CAD}
\end{figure}

The TF (see Figure~\ref{fig:TF_CAD}), as mentioned previously, is located at the entrance aperture of the instrument. It has a physical aperture of 154~mm and a clear aperture of 151~mm. The substrate for the TF is 10~mm thick UV-grade fused Silica (Corning 7980 1A) with surface micro-roughness of 1.5~nm RMS. 

The outer surface, directly facing the Sun, of TF substrate has a  custom designed multi-layer coating of Chromium (\rm{Cr}), Aluminum (\rm{Al}) and Silicon dioxide (\rm{SiO$_{2}$}). The coating deposition was done by Electron-beam physical vapor deposition process using a custom-developed handling and cleaning recipe to ensure no pinholes in the coating. The substrate was chemically cleaned to remove residual particles and organic contamination before loading into the coating chamber. Once inside the chamber, the substrate was Ion-cleaned with Argon for eight minutes. The 300~{\AA} thick \rm{Cr} layer was then deposited on the substrate in two stages during which the thickness of the coating was monitored by a Quartz Crystal Micro-balance (QCM). After half of the \rm{Cr} thickness was deposited, the filter was removed from the chamber. The \rm{Cr} surface was cleaned with Acetone and optical wipes to remove any residual particles. Thereafter, the substrate was loaded in the chamber and ion-cleaned before depositing the remaining thickness of \rm{Cr}. The substrate with full thickness of \rm{Cr} was removed from the chamber and cleaned using the process discussed above. Finally, the 350~{\AA} thick layer of Al and 350~{\AA} thick SiO$_{2}$ capping layers were deposited in a single run of the coating chamber (\cite{ghosh_tfa}) (see Table~\ref{tab:tfdesign}).

\begin{table}[!ht]
\caption{Details of the coating layers for the thermal filter (\cite{ghosh_tfa})}.\label{tab:tfdesign} 
 \begin{tabular}{c c c}
 \hline
 Layer No. 			& Material  				& Thickness\\
 \hline
 Substrate 			& UV-Fused Silica 			& 10 mm \\
 					& (Corning 7980 1A) 	    & \\
 Layer 1 			& Chromium 					& 300~{\AA}   \\
 Layer 2 			& Aluminum   	            & 350~{\AA}  \\
                     & SiO$_{2}$                 & 350~{\AA}\\
\hline
\end{tabular}
\end{table}
\begin{figure}[h]
\centering
\includegraphics[width=0.8\textwidth]{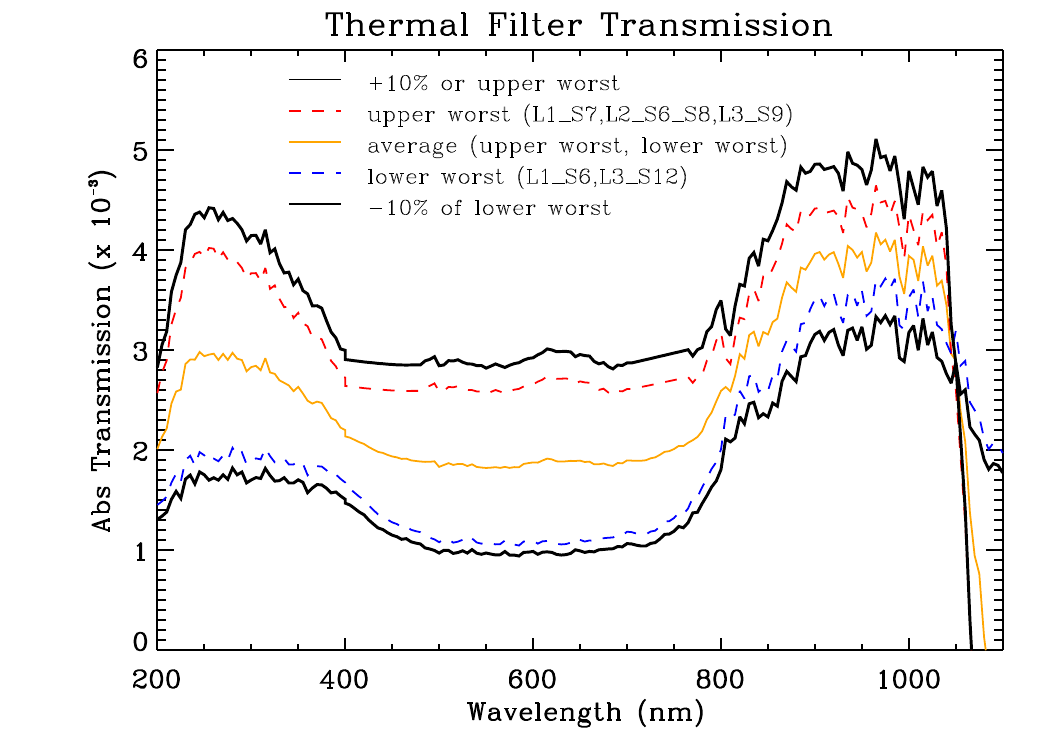}
\caption{Absolute transmission profile of the thermal filter as a function of wavelength. The solid yellow curve shows a mean transmission profile based on measurements over 30 representative samples. The red and blue dashed curves enclose the variations in the transmission values obtained for several coating attempts. The black solid curves include a $\pm$10\% margin on these extreme profiles.}\label{fig:TF_Trans}
\end{figure}

The TF is seated in the support ring (Ti-6Al-4V) and glued at three points using 3M™ Scotch-Weld™ Epoxy Adhesive EC-2216 B/A and secured by a lock ring (Ti-6Al-4V). The support ring is mounted on the three-point stand which is attached to the optical bench. Glass Fibre Reinforced Polymer (GFRP) spacers are used for thermal insulation between the bench and the TF stand to provide a high thermal resistance path between the filter and the optical bench. 

The TF coating was designed to meet two requirements for the instrument; first, it prevents the solar flux in the visible range from entering into the optical cavity. Second, it reduces the incoming solar flux within the operational wavelength between 200{--}400~nm. While the first is based on the science requirement of observations being made in specific passbands, the second is crucial in maintaining the CCD linearity and desired photometric performance. The transmission curve shown in Figure~\ref{fig:TF_Trans}, suggests that there is more IR flux entering the system than in the wavelength band of interest. This was due to the manufacturing limitations of the thermal filter to achieve enough photons to obtain the desired signal to noise ratio in wavelength band of interest. However, this issue is mitigated using the science filters that tightly selects the desired wavelength, preventing the incoming IR flux from reaching to the CCD.

The design and development of the TF was carried out in collaboration with Luma Optics Ltd\footnote{https://luma-optics-pvt-ltd.business.site} in Mumbai, India that included several design iterations and prototyping over a period of five years. The two major challenges were to decide upon the selection of substrate and coating. While the former is important for minimizing the possibility of pinholes, the latter is to ensure durability to withstand the cleaning process, as well as contamination-free assembly for high throughput in the NUV region and space qualification.

In Figure~\ref{fig:TF_Trans}, with a solid black line, we plot the mean transmission profile of a set of 30 TF representative samples measured using a Perkin Elmer Lambda 9 spectrophotometer. To account for the non-uniformity of transmissions due to imperfect coatings over multiple trials, we consider the measured maximum and minimum transmissions at all wavelength bins among each of these 30 representative samples. The resultant transmission profiles for these 30 representative samples are indicated by the solid blue and solid red curves. A further $\pm$10\% error on these measured profiles (shown as black dashed curve) still meets the performance requirement of the instrument. As it is evident from the curve, in the wavelength band of interest (200{--}400~nm) the TF allows the transmission to be $\approx$~0.2{--}0.3\%. For the visible the transmission falls to $\approx$~0.1\%. 

Furthermore, we note here that the TF has high reflectance ($>$70\%) on the outer surface, in the NUV and above wavelength domains. This ensures that most of the VIS and IR radiation is reflected and does not lead to excessive heating and deformation of the filter. The total heat load in the optical cavity of the telescope is limited to $\approx$~40~mW (for a detailed information see \cite{thermal, thermal2}). Detailed thermal and thermo-elastic deformation studies of the assembly were conducted to verify that the filter's thermal deformation (due to the heat loads) does not significantly affect the optical performance of the system. 

To ensure these filters survive the launch and orbital environment, rigorous space qualification and environmental testing of several TF representative samples were conducted at the Laboratory for Electro-Optics Systems (LEOS) and the U. R. Rao Satellite Center (URSC). Furthermore, to ensure that the TF survives the radiation environment during transit and five-year mission at L1, several TF representative samples were subjected to gamma radiation testing at URSC facility, followed by a proton radiation test at the BARC-TIFR Pelletron facility. For further details on the development, testing, and qualification of the TF \citep[see][]{thermal, thermal2}.

\subsection{Primary and Secondary Mirror Assemblies} \label{S-Mirrors}
The primary mirror (PM) is an off-axis concave-hyperbola with a 141~mm clear aperture and 159~mm off-axis distance. The Radius of Curvature (RoC) and Conic Constant (CC) of the PM are {--}1404.67~mm and {--}1.0255, respectively. The secondary mirror (SM) is an off-axis convex-hyperbola with 44~mm clear aperture with the off-axis distance of 40~mm. The RoC and CC for the secondary mirror are {--}438.25~mm and {--}2.4867, respectively. The mirrors are fabricated on Zerodur substrates, which has a low thermal expansion coefficient that minimizes the optical distortion due to thermal effects. Both mirrors are coated with custom designed protective aluminum coating with $>$86\% reflectivity on average in the 200{--}400~nm bandpass. The substrates are  polished to a high precision to achieve a low surface micro-roughness, better than 10~{\AA} RMS for the primary and 7~{\AA} for the secondary, to reduce scattering. 

\begin{figure}
\centering
\includegraphics[trim=2cm 2cm 2cm 2cm, angle=270, width=0.9\textwidth]{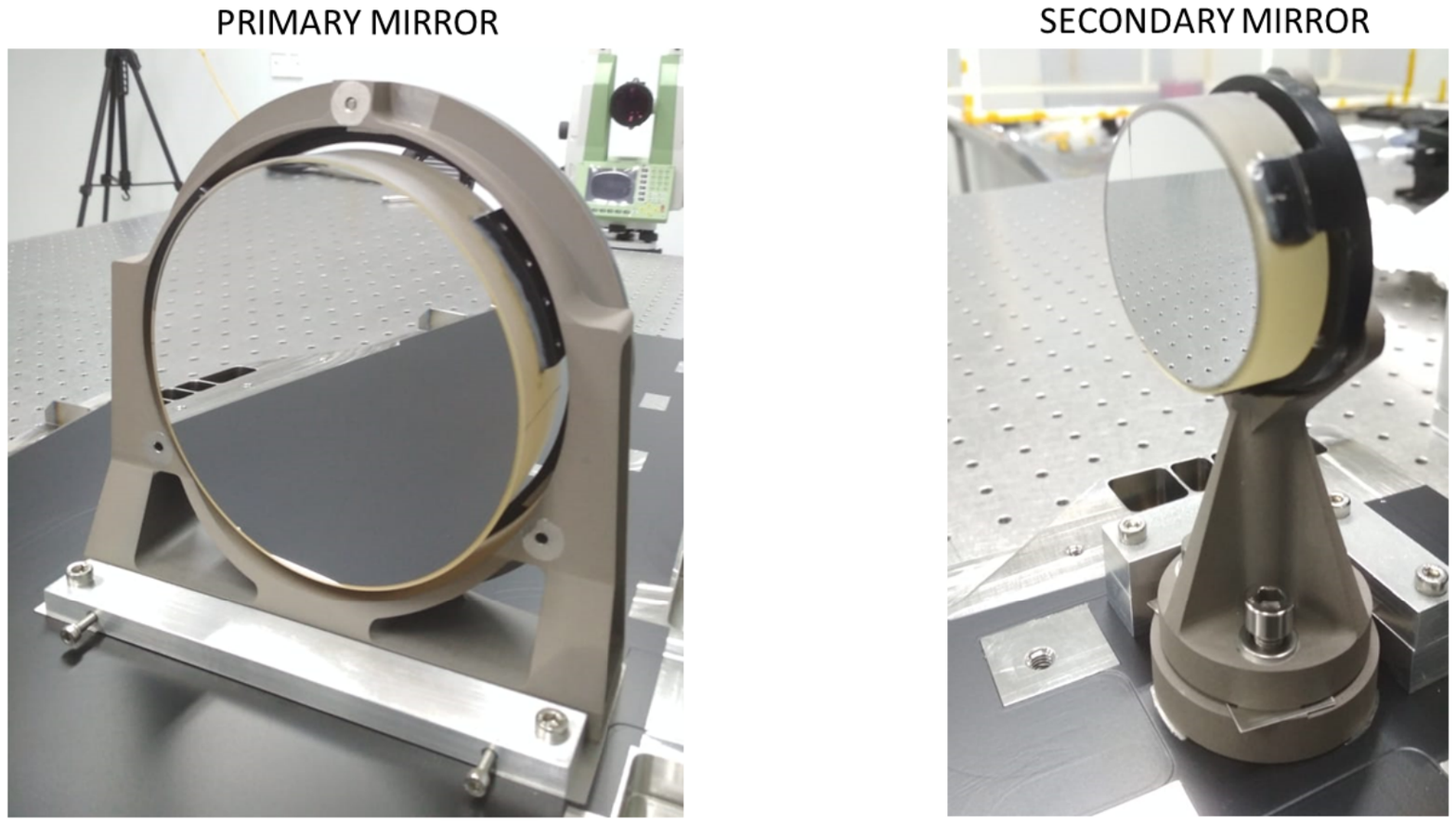}
\caption{Flight model of the Primary (left) and the secondary (right) mirror assemblies.} \label{fig:suit_pm_sm}
\end{figure}
\begin{figure}
\centering
\includegraphics[width=0.9\textwidth]{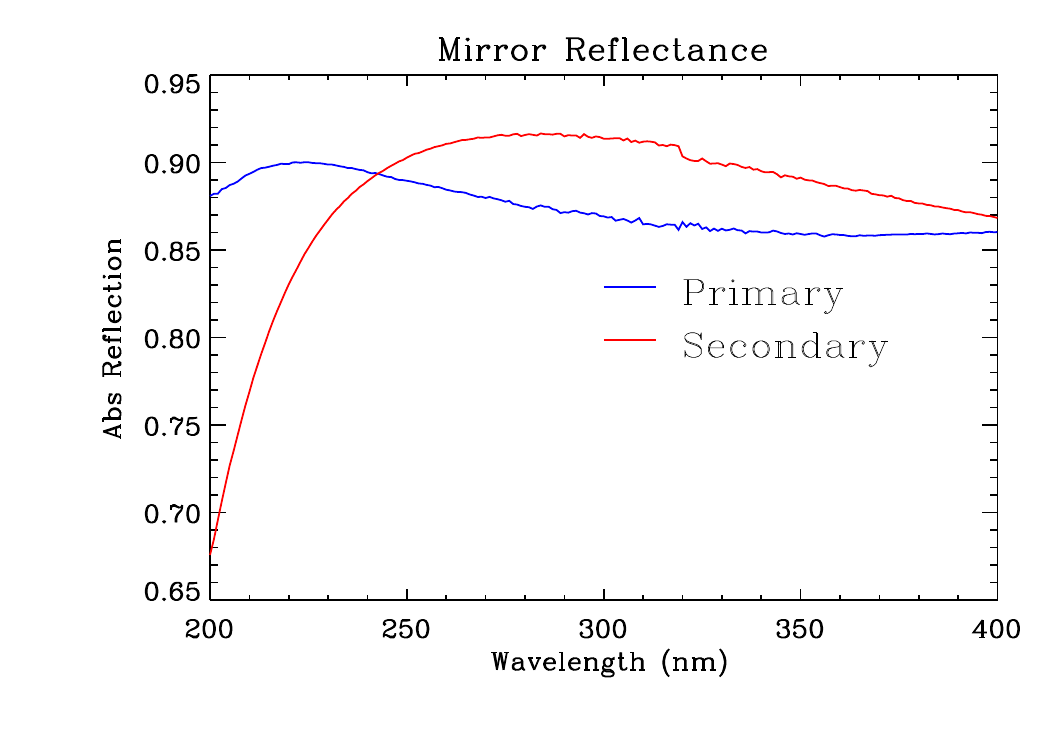}
\caption{Absolute reflectance of primary and secondary mirrors. } \label{fig:mirrors}
\end{figure}

The picture of the flight models of the primary and secondary mirror assemblies are shown in Figure\ref{fig:suit_pm_sm}. Both the mirrors and their mounts were designed and developed by the LEOS, an ISRO centre in Bengaluru. The mirrors are bonded to low-stress flexures on invar rings. The mirror and invar ring assemblies are mounted on stands made of Titanium alloy (Ti-6Al-4V). The mirrors, with their respective ring and mount along with the ring and the stand, are referred to as Primary Mirror Assembly and Secondary Mirror Assembly. The mirror assemblies are mounted on the optical bench. The optical axis height from the bench is 100~mm. The SM stand has an additional spacer plate, which is machined to compensate for adjusting optical axis height and for correcting its tip-tilt for optical alignment. The structural design of the mirror assemblies has been done to ensure that stresses on the optics due to launch loads are well within the safe limit ($<$~5~MPa) and that the deformation of optical surfaces due to mounting and thermal loads does not affect the optical performance. Figure~\ref{fig:mirrors} plots the absolute reflectance of both the primary and secondary mirrors.

\subsection{LED Calibration Unit}\label{S-LED}
\begin{figure}[h!]
\centering
\includegraphics[width=0.7\textwidth]{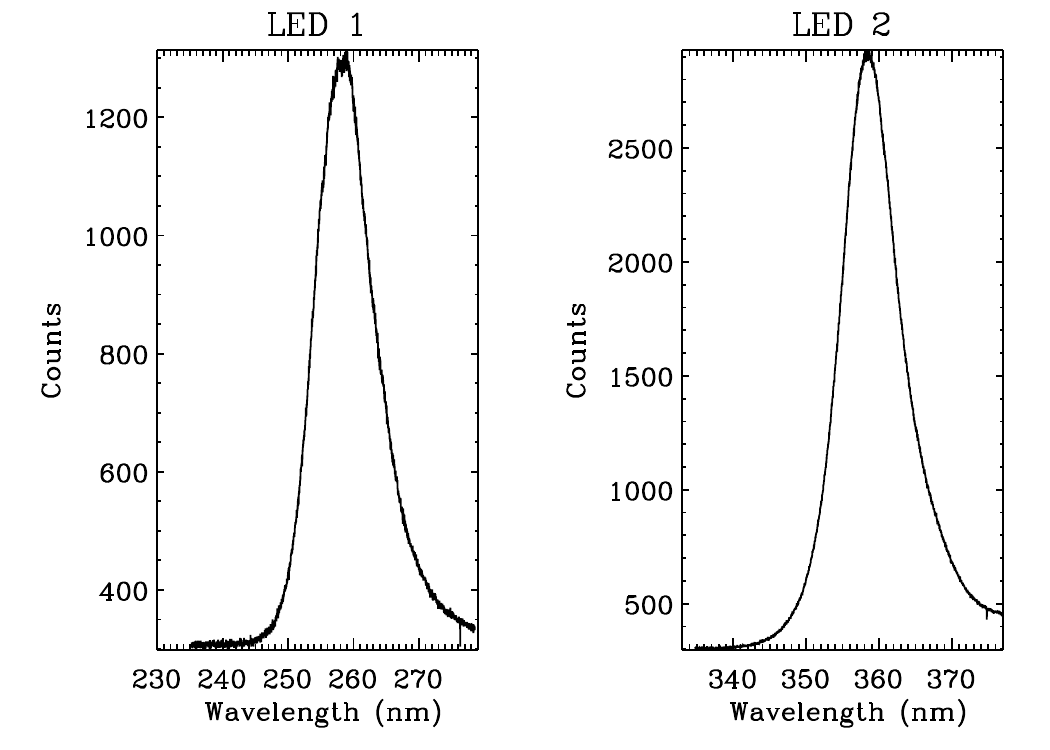}
\caption{Spectrum for LED1 (left) and LED2 (right) }\label{fig:LED_spec}
\end{figure}

The LED calibration unit is located between the shutter and the filter wheel. It is designed to provide uniform detector illumination for calibration and long-term degradation studies. The LED calibration unit consists of sixteen LEDs (eight each for peak wavelengths at 258 nm and 356 nm) mounted on a Printed Circuit Board (PCB) in a circular configuration around a cutout for the beam from the secondary. The eight LEDs of a particular wavelength are grouped in two sets of four (main and redundant) such that each set creates a uniform illumination on the detector and can be turned on and off independently. The calibration unit is used in the conjugation of broad-band filters and complimentary filters that transmit at the central wavelengths of the LEDs. The specifications for the LEDs are given in Table~\ref{led} and their spectrum is shown in Figure~\ref{fig:LED_spec}.

The SUIT LEDs are commercial of-the-shelf components developed by Sensor Electronic Technology, Inc. (SETi). These have been procured, screened and space-qualified at URSC. The space qualification program for these LEDs included a suite of mechanical, electrical, environmental (thermal cycling, thermal soak etc.) and radiation tests (gamma and proton radiation) as per ISRO's standards for Electrical, Electronic and Electromechanical (EEE) parts. The flight version is screened and selected from the same lot that has been used for non-destructive space qualification tests. 

\begin{table*}[!ht]
\caption{Characteristics for on board calibrations LEDs} \label{led}
\begin{tabular}{ccc}
\hline
Characteristic              & LED~1  			& LED~2 \\
\hline
Peak Wavelength (nm)        & 258    		& 356\\
FWHM (nm) 			           & 11 		& 11  \\ 
Peak power (mW) 	       & 180			    & 112\\
Peak Current (mA)          & 30                & 30 \\
Forward Voltage (V)	        & 6 		& 3.75 \\
\hline
\end{tabular}
\end{table*}   

\subsection{Filters and Filter Wheels}\label{S-Science_filters}
\begin{figure}
\centering
\includegraphics[width=0.8\textwidth]{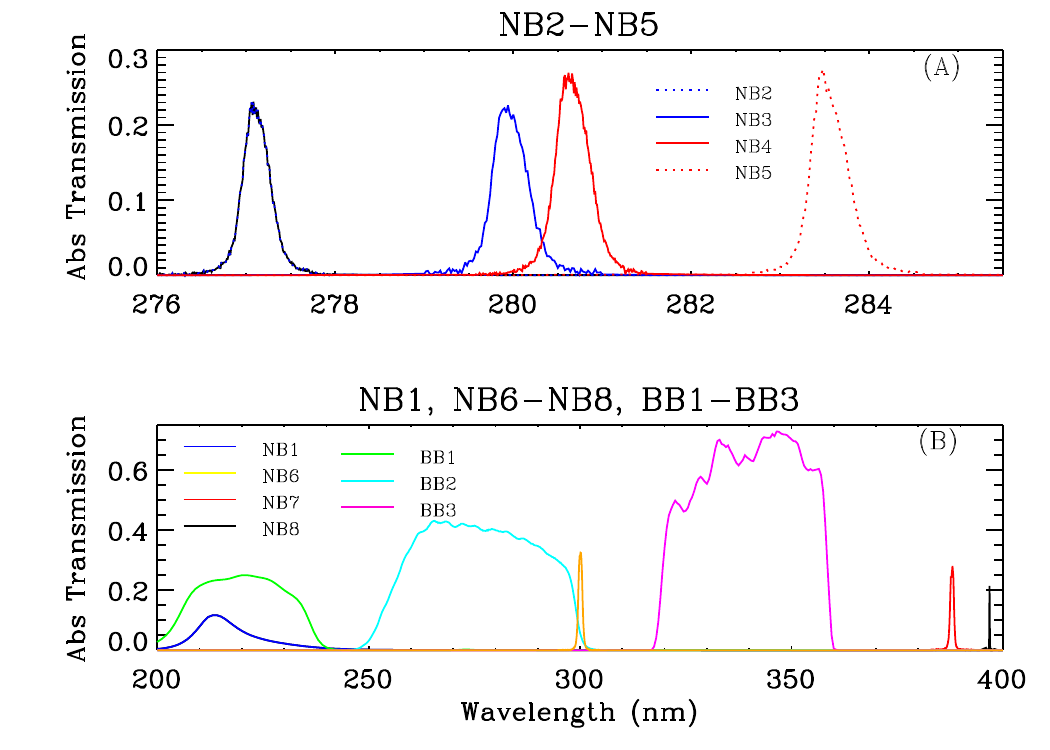}
\includegraphics[width=0.8\textwidth]{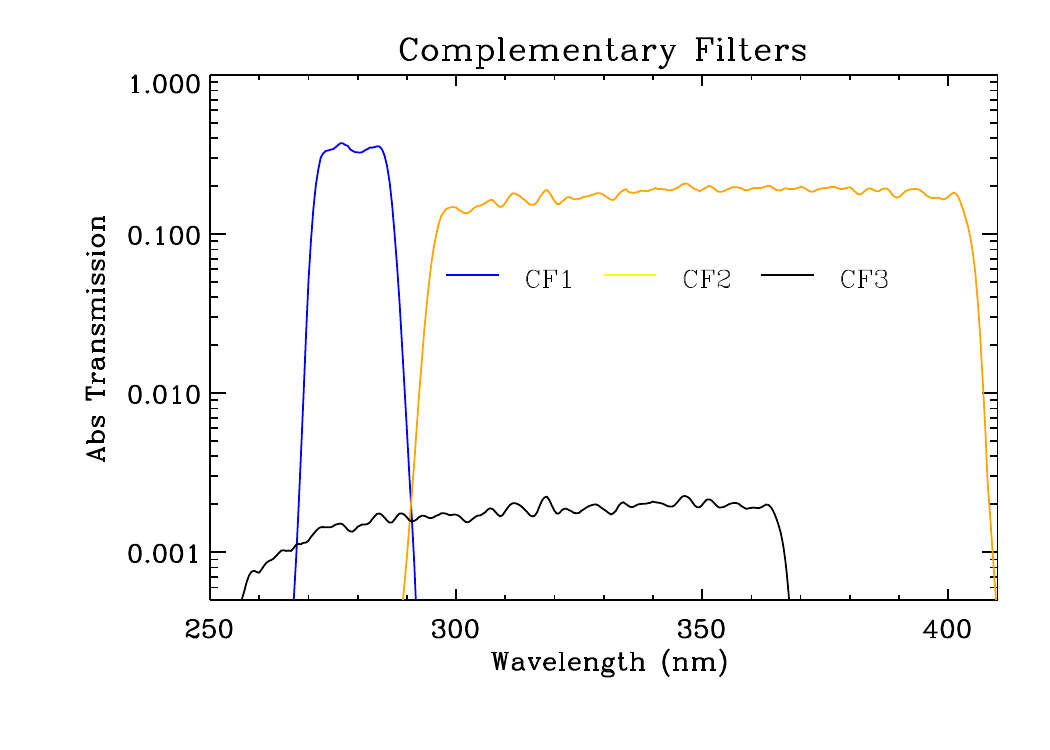}
\caption{Absolute transmission profiles of the eleven SUIT Science filters (top and middle) and complementary filters (bottom) as a function of wavelength.}\label{fig:science_filters} 
\end{figure}

The light reflected from the secondary mirror passes through two filters before reaching the CCD. For some science filters, e.g., NB02, NB03, NB04, NB05, NB06, and NB07, we needed to design separate complementary filters, while NB01 and BB01 combine with BB01 and NB08 combines with another NB08. In total {\suit} carries 16 filters mounted on two different filter wheels, each having 8 slots. The details of the eleven science filters are noted in the first three columns of Table~\ref{sc_comb_fil}. The fourth column lists the combination filters corresponding to each science filter. The transmission profiles of all the filters are shown in Figure~\ref{fig:science_filters}. All these filters have been characterised for their spectral response in the lab before finally mounting into the instrument \citep[][]{jj_filter}. All the science filters were custom manufactured by Materion Corp based on the provided requirements.

As mentioned earlier, each filter wheel has eight slots with constraints related to the Mg-filters and that a filter on one wheel can be combined at most with five filters on the other wheel. Furthermore, the complementary filter assignment and arrangement of all filters on these wheels have been done targeting optimized cadence for normal operational modes of the instrument. 
The combined transmission of the TF, a given science filter, and its complementary filter further ensure that unwanted leak beyond the passband is minimized with the desired cut-down within 200{--}400~nm. The effective transmission profiles, accounting for a given science filter, its combination filter, and other optical components in the ray path, meet SNR requirements within the nominal exposure range.

\begin{figure}
\centering
\includegraphics[trim=2cm 2cm 2cm 2cm,angle=270,width=0.6\textwidth]{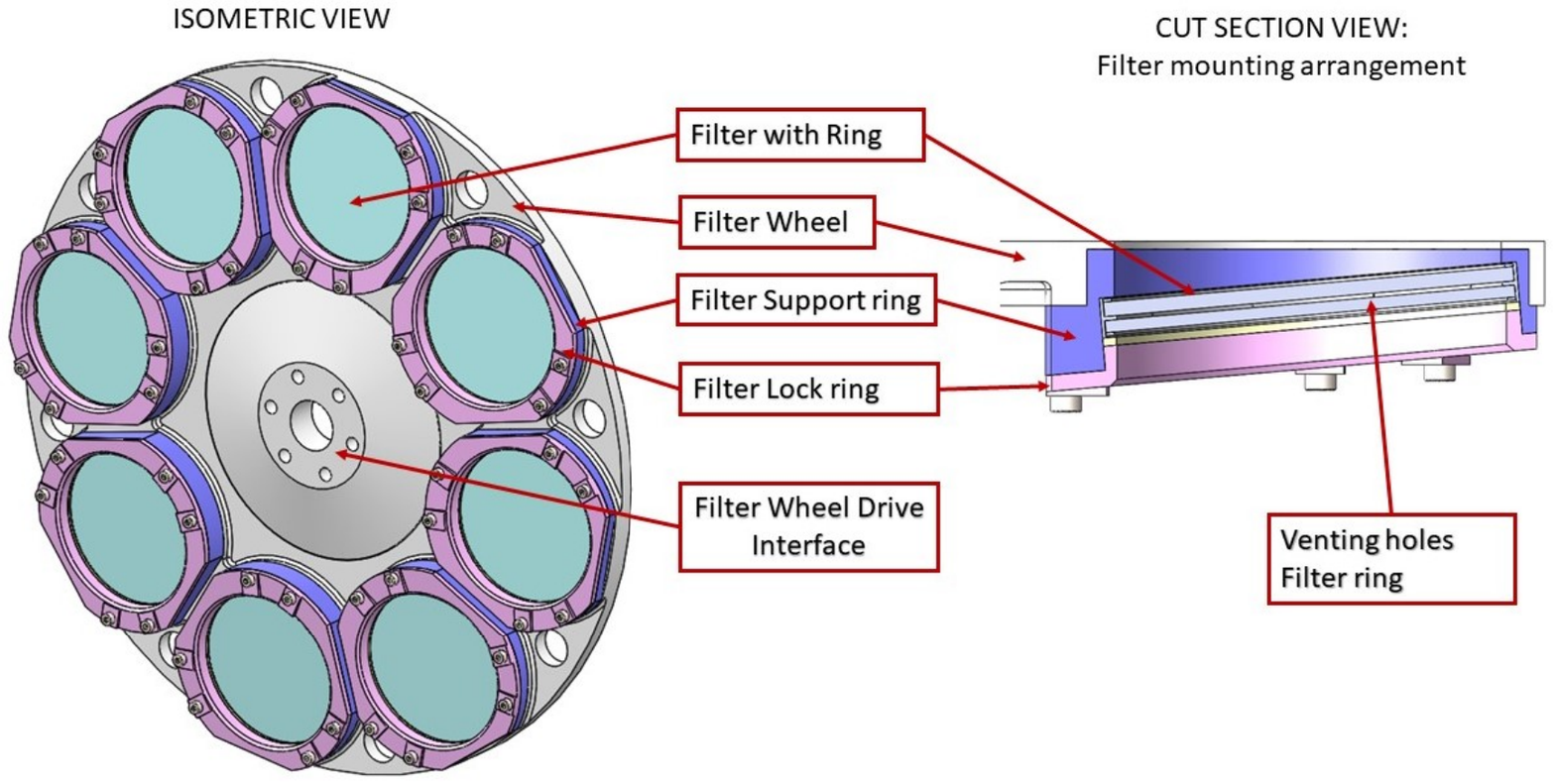}
\includegraphics[trim=2cm 1.5cm 0cm 3.5cm,angle=270,width=0.6\textwidth]{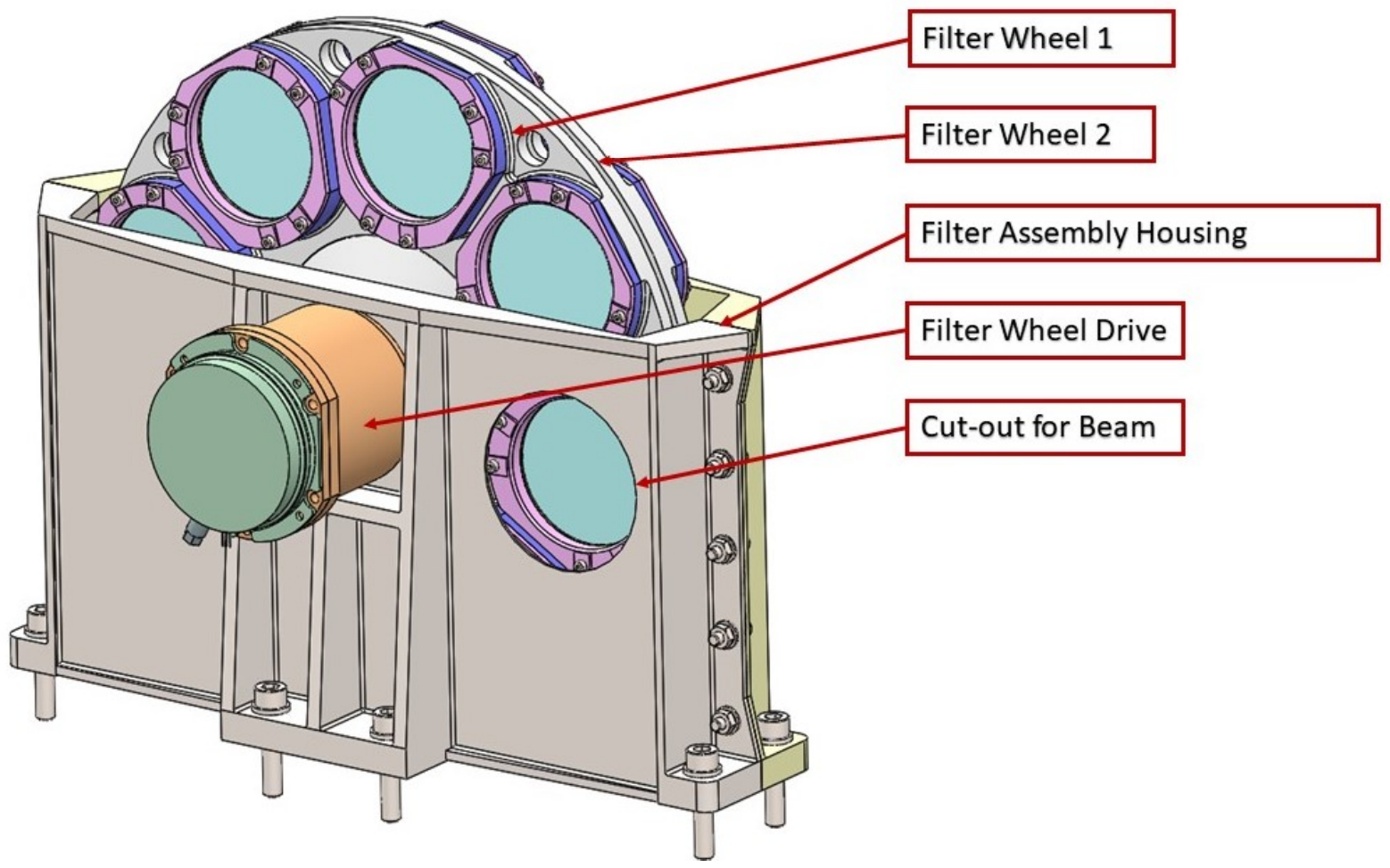}
\includegraphics[width=0.4\textwidth]{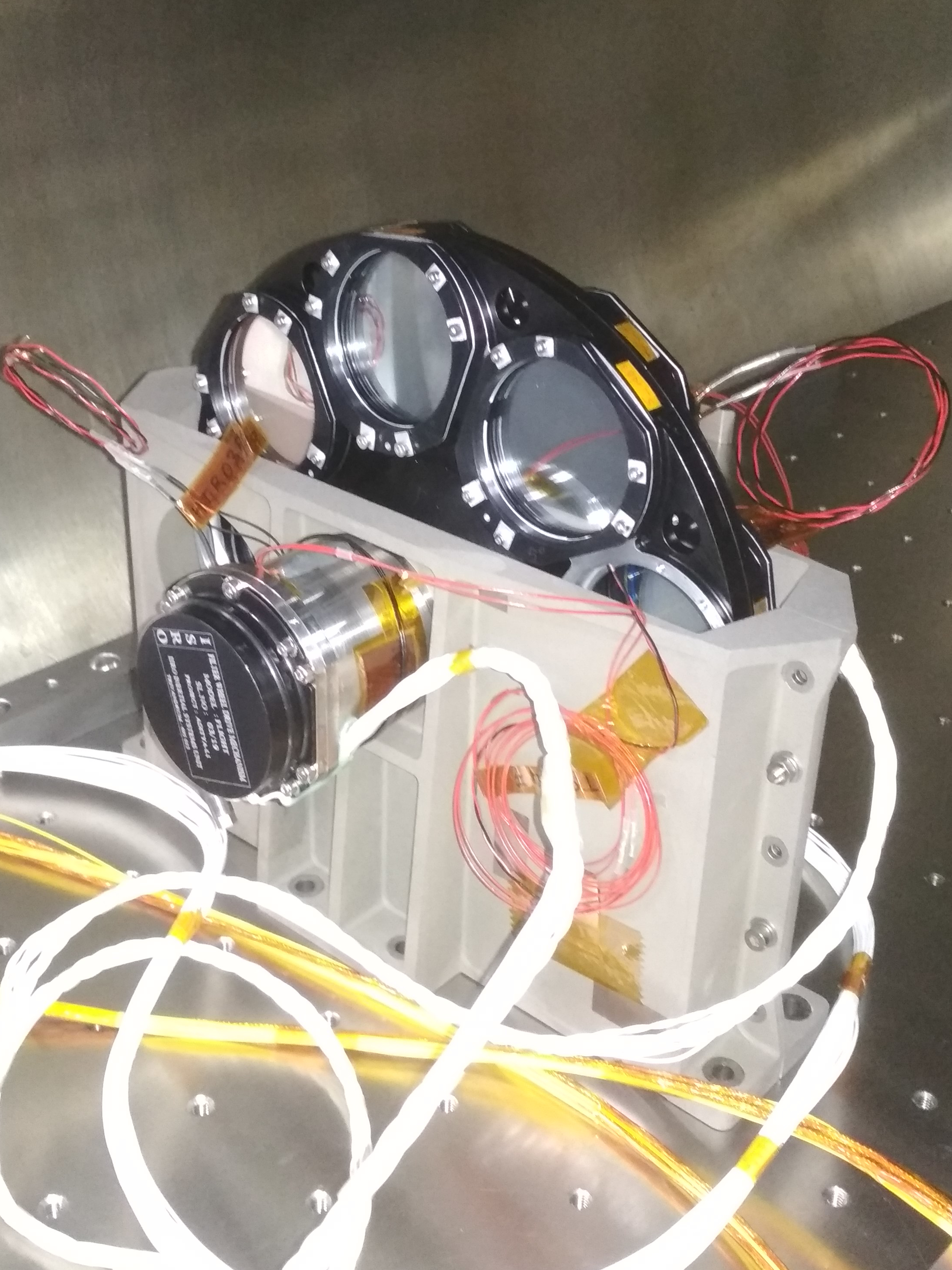}
\caption{Top: Isometric CAD view of the filter wheel (left) and cut section of the filter mounting arrangement with the tapered support ring to control the filter tilt with respect to the beam (right). Mid: CAD model of the filter wheel assembly. Bottom: Assembled flight model of SUIT Filter Wheel Assembly.} 
\label{fig:FW_Wheel}    
\end{figure}

Figure~\ref{fig:FW_Wheel} (top panel) shows the CAD model of one of the filter wheels with filters mounted in all its slots (left) and the cut section view of the filter mounting arrangements. The filter wheel and the holder rings are made of Aluminum alloy (Al 6061-T6). The assembly of individual filters, comprising the filter, the support ring, the locking ring, and the spacer, has been designed to ensure that the mechanical and thermal stresses on the filters during launch and operations are within the safe limit. To minimize the ghosting effects, the filters are tilted with respect to the beam. The tilt for the filters is incorporated by the tapered design of the support ring. 

The CAD model of the complete filter wheel assembly is shown in the bottom panel of Figure~\ref{fig:FW_Wheel}. The two wheels and drive mechanism are mounted on a common mechanical housing to allow for better co-alignment of the wheels and reduce the overall mass. Two independent stepper motors with an in-built hall-effect position sensor drive the two filter wheels. The desired combination of filters in the beam path is achieved by moving both filter wheels independently.

\subsection{Field Corrector Lens} \label{S-Lens}

The field corrector lens is a positive meniscus lens optimized to improve the system's imaging performance for a wide field of view. The lens is made of UV-grade fused Silica (Corning 7980 AA) and has aperture and thickness of 49~mm and 6~mm, respectively. The radii of curvature for the two surfaces are 32.971~mm (front) and 28.0369~mm (back). To minimize the scattering, the lens was fabricated with a low inclusion and a high homogeneity substrate, and the optical surfaces were polished the optical surfaces to 1.5~nm RMS micro-roughness to minimize the scattering. The lens has a custom multi-layer anti-reflection (AR) coating consisting of a total of eight alternative layers of Hafnium Dioxide (HfO$\textsubscript{2}$) and Silicon Dioxide (SiO$\textsubscript{2}$). It was designed to minimize unavoidable ghost artifacts created due to the proximity of the lens to the detector. The custom coating was developed and space-qualified by IUCAA in collaboration with LEOS and Luma Optics Ltd. 
\begin{figure}[h!]
\centering
\includegraphics[width=0.25\textwidth]{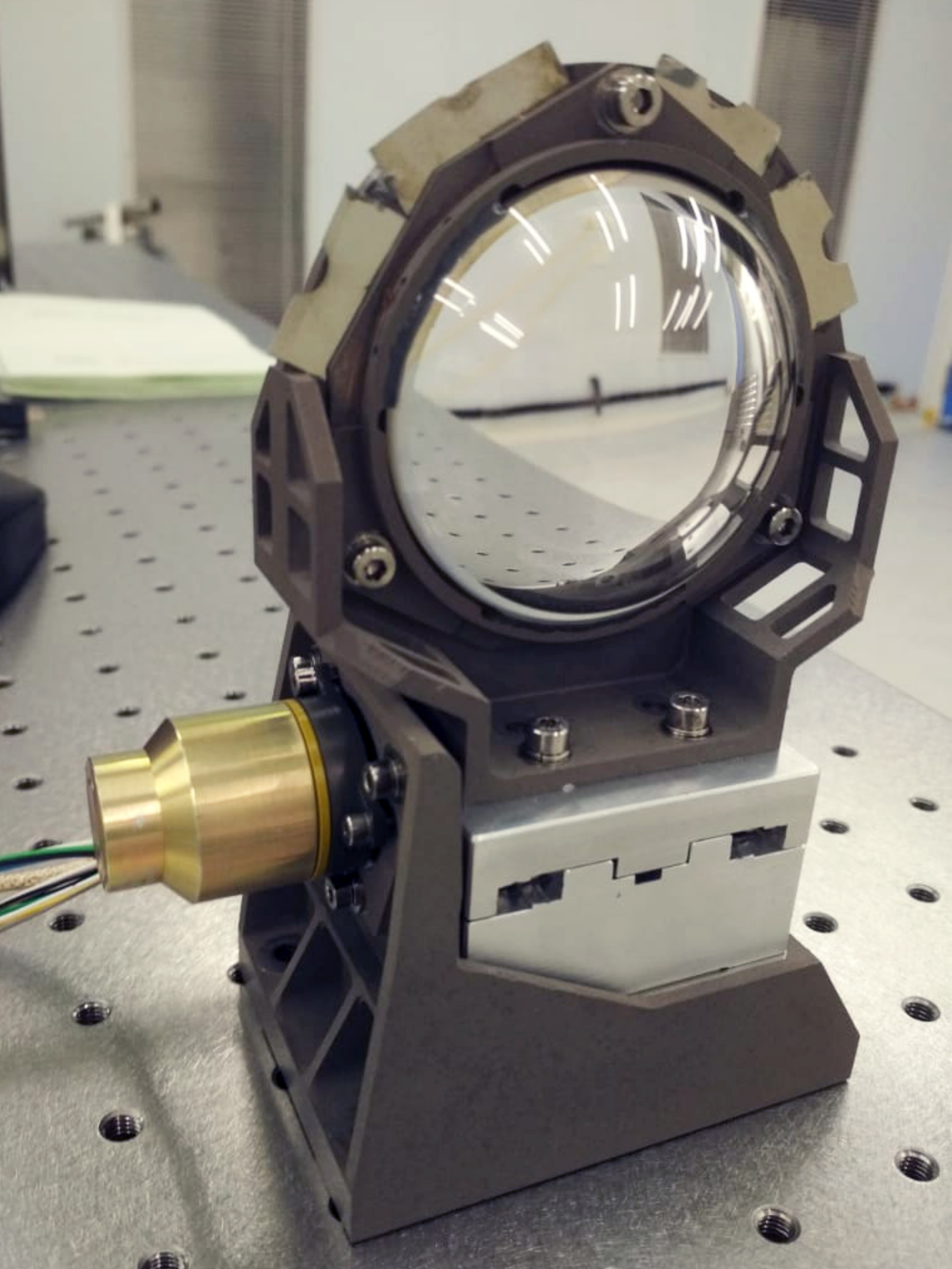}
\includegraphics[trim=0.5cm 1cm 1cm 0cm,width=0.5\textwidth]{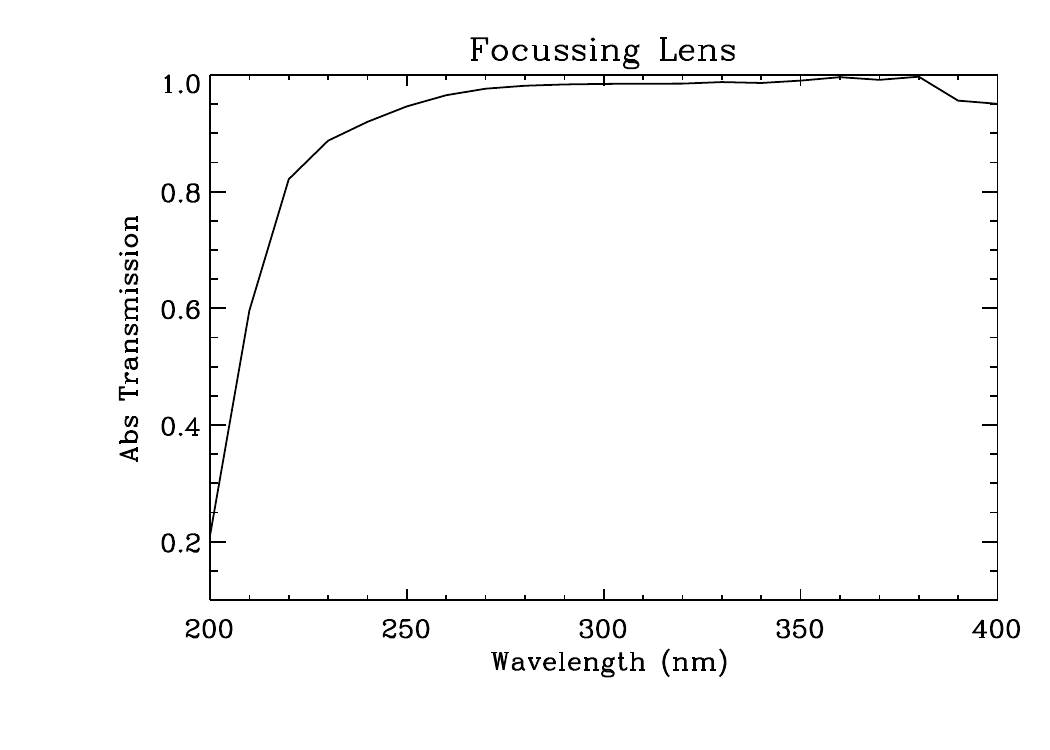}
\caption{Left: Flight model of field corrector lens assembly. Right: Absolute transmission of the field corrector lens as a function of wavelength.}
\label{fig:lens}       
\end{figure}

Figure~\ref{fig:lens} (left panel), shows the flight model of the field corrector lens assembly with the lens mounting arrangement that comprises the lens mount and the adapter. The lens has been designed to compensate for linear thermal distortions of the optical system and bench to provide acceptable image quality performance over a temperature range of 20~$\pm$3~$^{\circ}$C . The compensation for optimal focus can be accomplished by linearly moving the lens along the optical axis, with the help of a Piezo motor. The right panel of Figure~\ref{fig:lens} show the absolute transmission of the field corrector lens as a function of wavelength.

\subsection{Detector and Detector Housing Assembly}\label{S-Detector_Assembly}

\begin{figure}[h!]
\centering
\includegraphics[width=0.48\textwidth]{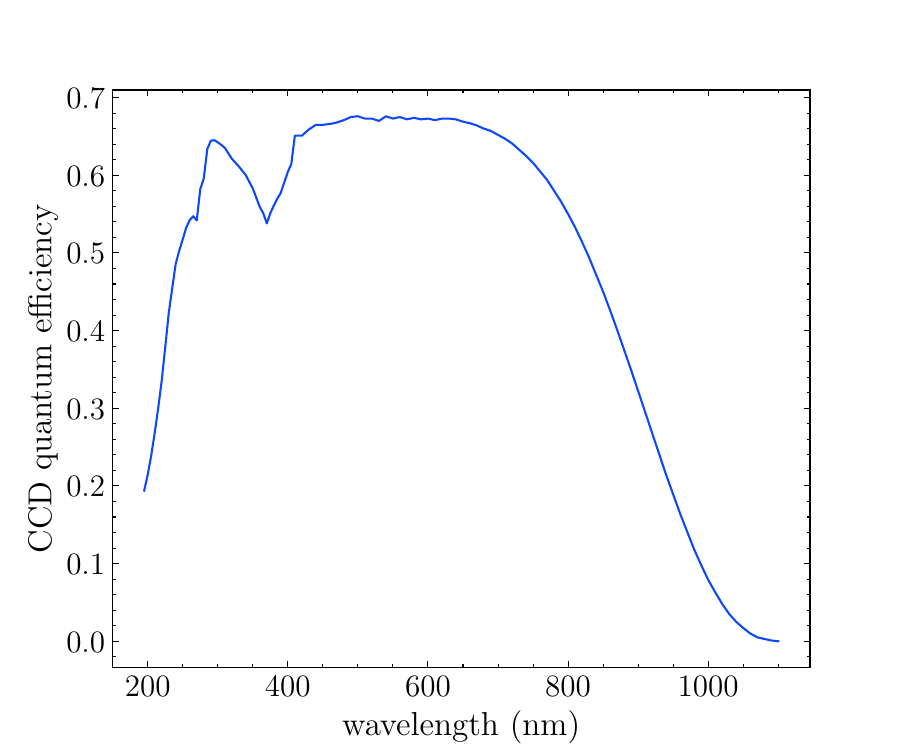}
\includegraphics[trim={1cm 1cm 1cm 1cm},clip,width=0.5\textwidth]{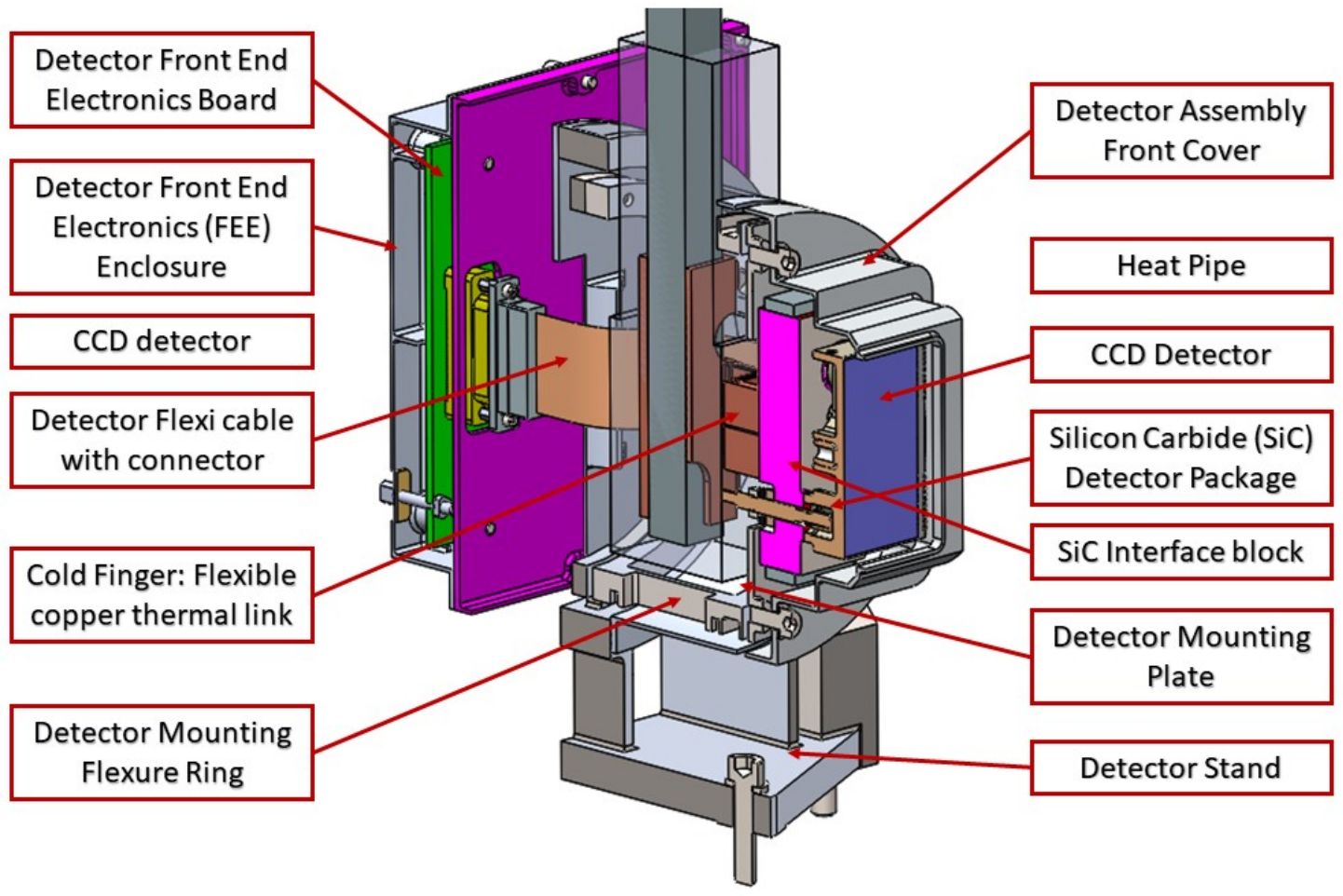}
\includegraphics[width=0.7\textwidth]{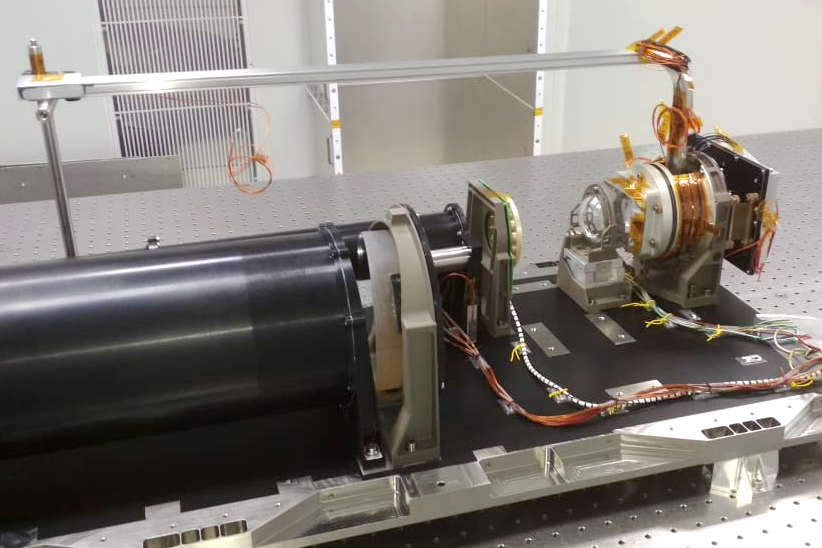}
\caption{Top left: Quantum efficiency curve for CCD 272-84 with custom anti-reflection coating. Top right: A CAD representation (sectioned view) of the detector Assembly showing the mounting scheme and cooling chain for the detector. Bottom: Detector Assembly and heat pipe mounted on SUIT.} \label{fig:CCDQE} 
\end{figure}

The detector for SUIT is a custom-built, UV-enhanced, back-thinned, line-transfer Charge-Coupled-Device (CCD), CCD272-84 manufactured by Teledyne e2v (UK) Ltd. The CCD has 4096~$\times$~4096 pixels with a pixel size of 12~$\mu$m. It has four readout amplifiers at the four corners that can be used to read the four 2048~$\times$~2048 quadrants simultaneously. The CCD has the anti-reflection coating of Hafnium Dioxide, HfO\textsubscript{2}, which is optimized to enhance the quantum efficiency (QE) in the wavelength range of interest and also to minimize the ghosting. The QE curve of the CCD in 200-1100~nm range is shown in the top left panel of Figure\ref{fig:CCDQE}.

\begin{figure}[h!]
\centering
\includegraphics[trim=1.5cm 0cm 0.5cm 0cm, angle=270,width=0.7\textwidth]{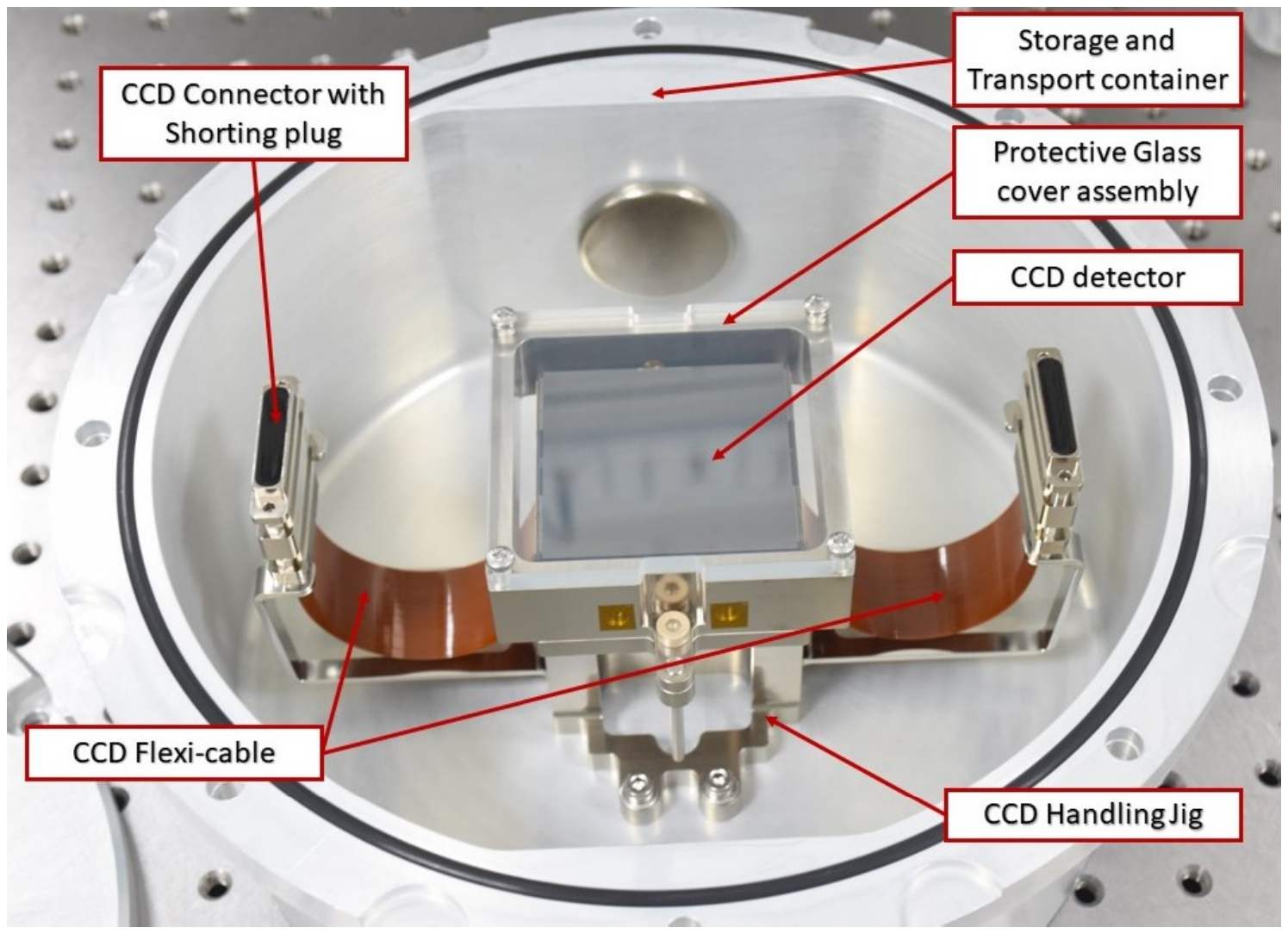}
\caption{Image of the flight model CCD in the transport container recorded during post-delivery inspection.}\label{fig:CCDFM1}       
\end{figure}
\begin{table*}[!ht]
\caption{Characteristics of the Teledyne e2v CCD272-84}.\label{tab:CCDQE}
\begin{tabular}{lr}
\hline
Parameter 					& Value 									\\
\hline
Image Area Format 			& 4096 $\times$ 4096 pixels 				\\
Image Pixel 				& 12 $\mu$m         					    \\
Operation Mode 				& Non-inverted               				\\
Full-well capacity 			& 190,000 e-  					\\
Pixel Readout frequency 	& 280 kHz (Nominal)							\\
Operating Temperature 		& {--}55 $^{\circ}$C						\\
Read Noise 					& $\approx$~4 e- for 280 kHz readout			\\
Dark Noise 					& 4 e-/pixel/sec @-55$^{\circ}$C			\\
AR Coating 					& Hafnium Dioxide (HfO$_{2}$) 				\\
Number of Outputs 			& 4 real, 4 dummy 							\\
Electrical Interface 		& flexi cables each with  					\\
							& a 37-way micro-D							\\
     \hline  
\end{tabular}
\end{table*}

The CAD model of a cross-section of the assembly is shown in the top right panel of Figure~\ref{fig:CCDQE}. The CCD is mounted on a Silicon Carbide (SiC) package that provides the mechanical interface with a Titanium alloy mount assembly designed to maximize the heat path between the detector and the optical bench. The mount has covers to block stray light and shield the cold detector from ambient IR and molecular contamination. The mount provides insulated support to the CCD flexi-cable that connects to the front-end electronics (FEE). The FEE enclosure attached to the instrument back panel is connected to the mount via a flexible support link made of thin titanium alloy blades to minimize the differential vibration between the detector and the FEE enclosure. 

The SiC block also provides the thermal interface for the detector and is coupled to the passive radiative cooling assembly by a flexible copper cold finger. It is insulated from the mount assembly by GFRP pads, which have high thermal resistance and low thermal expansion. The cooling chain comprises a flexible copper cold finger connected to the ethane heat pipe. Additional thermal insulation of cold finger and heat pipe is accomplished by Multi Layer Insulation (MLI) wrapping.
The detector is passively cooled to a temperature of \(-\) 55 $^{\circ}$C through a cold finger that is connected to a passive radiator. Cooling to \(-\) 55 $^{\circ}$C is required to achieve a dark current value below 4 e~pix$^{-1}$~s$^{-1}$, which is similar to the CCD readout noise for the nominal readout frequency of 280 kHz \citep[][]{Varma_ccd}. The CCD is operated in the non-inverted mode to optimize for the FWC ($\approx$~190~ke~pix$^{-1}$). A summary of the operational and performance characteristics of the CCD is given in Table~\ref{tab:CCDQE}.

The qualification and flight models of the CCD have been delivered by Teledyne after a 30-month-long building and testing phase. Figure~\ref{fig:CCDFM1} shows the image of the flight model for the CCD during post-delivery inspections. The flight model was selected from a custom-built lot of devices through a reduced lot acceptance test (LAT), which included electrical, mechanical, and thermal tests, cosmetic inspections, and proton and gamma radiation testing. The reduced LAT was conducted based on full space qualification of similar devices (with similar architecture and fabrication processes) for VIS Imager \citep[][]{VisE_2012} on board the EUCLID \citep[][]{Euclid} of the European Space Agency (ESA).

\section{Mechanisms}\label{S-mechanisms}
\subsection{Entrance Door Mechanism} \label{S-Door}

\begin{figure}[h!]
\centering
\includegraphics[trim=2cm 2cm 0cm 2cm,angle=270, width=0.8\textwidth]{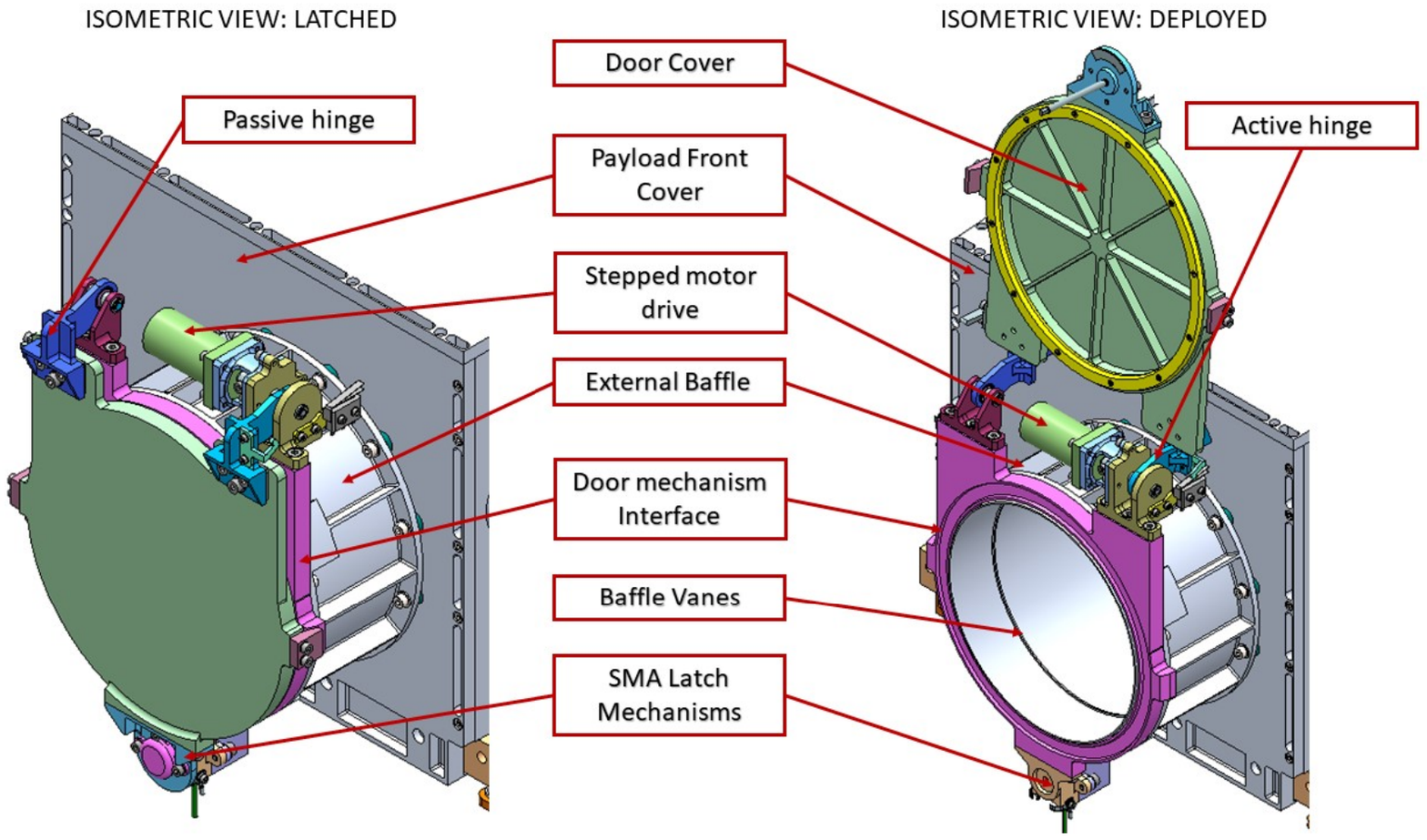}
\includegraphics[width=0.5\textwidth]{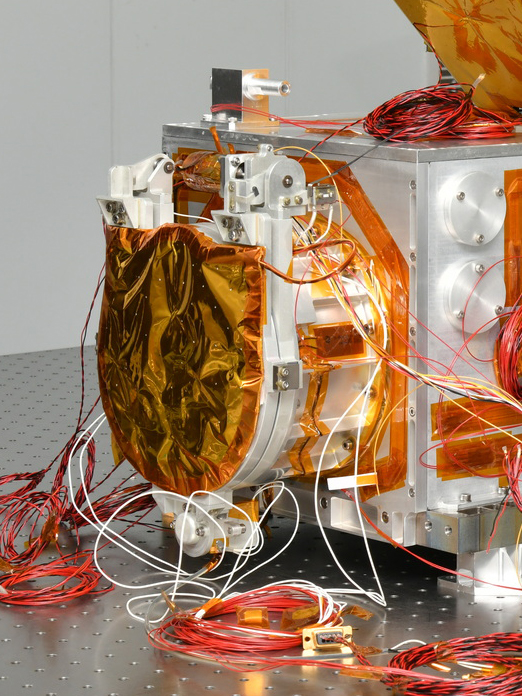}
\caption{A CAD representation of the multi-operational door mechanisms showing the stowed (top-left) and deployed (top-right) configurations. The door mechanism mounted on the external baffle of SUIT (bottom).}\label{fig:SUIT_Door}
\end{figure}

Figure \ref{fig:SUIT_Door} shows the CAD model of the entrance door mechanism assembly, which is designed and developed by the mechanisms group at UR Rao Satellite Center (URSC). It consists of a door mounted on the extended external baffle, a hold-down release actuator, and a stepper motor mechanism that drives the door at the two hinges. The stepper motor and door hinges (active and passive) have been designed for more than 500 operations, commensurate with the mission's nominal lifetime.

The door protects the optical cavity from contamination during ground operations, launch, and transit phases. The multi-operation capability of the stepper drive is designed to close the door before the S/C thruster fires during the periodic orbit correction maneuvers. This precautionary provision prevents thruster plumes from entering the instrument optical cavity.

During the launch and transit phase of the mission, the door will be latched in the closed position by a single-operation Frangi bolt actuator. The door was opened after a bake-out period to prevent degassed material from the spacecraft from entering the optical cavity. 

The deployment of the actuator and operation of the door mechanism are controlled directly by the S/C on board computer (OBC). The nominal operation of the stepper motor is done through ground commands, and the door can be fully deployed to an angle of 95\textsuperscript{$\circ$} to the optical axis. The pre-programmed on board commands are used for automated door operations when the S/C goes into safe mode or before thruster firing operations are performed.

\subsection{Filter Wheel Mechanisms} \label{S-Filter_Wheel}

The filter wheel assembly consists of two independent filter wheel mechanisms. Each mechanism contains a filter wheel, a drive motor, and a set of 8 Hall-effect positions sensors at 45\textsuperscript{$\circ$} corresponding to 8 filter slot locations. The two filter wheels are mounted on a common stand in a back-to-back configuration such that the motors point in opposite directions. The drive motors are hybrid stepper motors designed for two-phase bipolar operations. These stepper motors have 200 steps per revolution, i.e., 1.8\textsuperscript{$\circ$}/step, with the capability of bidirectional operation. The filter in the beam path can be changed in less than 1.5 seconds, including a 0.5-second settling time. The micro-stepping capability allows to achieve a position accuracy within $\pm$ 15 arc-minutes with repeatability. The drive profile has been designed to minimize the vibrations during the motor operations. Nevertheless, the satellite has a mechanism to simultaneously correct the reaction torque effect on the satellite. 

Motor bearings with low-outgassing inorganic lubricants have been specifically selected to minimize molecular contamination while meeting the lifetime requirement of roughly 5 million revolutions. To prevent oxygenation of the lubricants, the motor is kept under high-purity nitrogen-purged conditions during ground operations and testing. The motor has an inherent detent torque even without coil excitation. The detent torque and bearing friction are sufficient to hold the filter wheel in position during launch. 

The control loop for the filter wheel drive uses the Hall-effect sensors mounted on the drive motor shaft to determine the filter's position. A position knowledge within $\pm$~30~arc-minutes can be achieved based on the motor characteristics and the position sensor feedback. The current location of the filter wheel is sent to telemetry after each filter wheel operation. 

During nominal operations, the processing electronics provide the number of steps and direction (depending on the operational mode) to the Filter Wheel Drive Electronics (FWDE) that generates the drive profile and operates the drive. Once the desired position is achieved, the FWDE acknowledges the PE on the successful execution of the command. The FWDE also sends the command and positions the log to the ground directly through the S/C telemetry. 

The design and development of the drive mechanism have been done at ISRO Inertial Systems Unit (IISU), Thiruvananthapuram, and it is based on the design of the drive mechanism used in the Ultraviolet Imaging Telescope \citep[UVIT;][]{uvit} on board Astrosat \citep[][]{astrosat}. 

\subsection{Shutter Mechanism} \label{S-Shutter}
\begin{figure}[h!]
\centering
\includegraphics[width=0.5\textwidth]{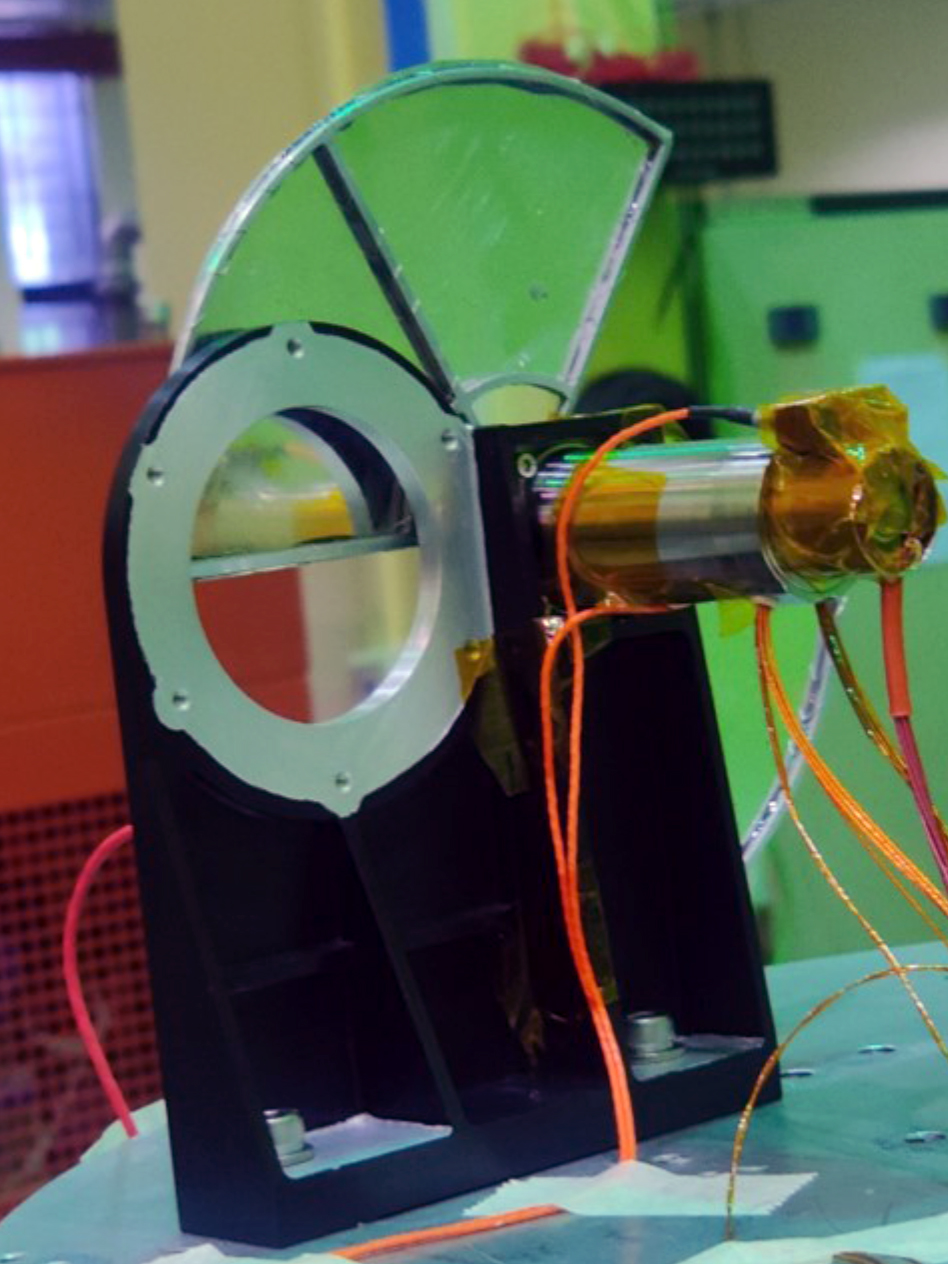}
\caption{Image of the flight model of the shutter mechanism assembly. The motor with the blade is co-mounted on the secondary baffle mount next to the filter wheel assembly.}
\label{fig:Shutter_mech} 
\end{figure}

The shutter mechanism is designed for exposure control. The shutter frame is made of a thin aluminum frame with a 50-micron thick vane made of stainless steel sheet glued on the frame. The whole vane is painted black with inorganic paint and is rotated by a stepper motor. The blade has a 156~mm diameter with two symmetrical openings of 56\textsuperscript{$\circ$} each. The blade is rotated 180 degrees for each exposure to sweep the light beam across either one of the openings. The shutter mechanism is mounted on a common mount with the secondary baffle, as shown in Figure~\ref{fig:Shutter_mech}.

For exposures less than $\approx$~300~ms, the blade is continuously rotated to sweep the beam with the blade opening. Changing the sweep velocity achieves the desired exposure duration on the CCD. Longer exposures are achieved by moving the blade and holding the opening in front of the detector. The shortest exposure that can be taken using this mechanism is $\approx$~100~ms and is limited by the motor's peak torque. The mechanism is designed and tested for 11 million rotations, required to take 22 million exposures over the 5$-$year mission lifetime.

The drive motor is a custom-built Phytron Physpace 2-Phase, hybrid, stepper motor that is capable of providing 18~mNm torque at a peak operation velocity of 160~rpm and a holding torque of 18~mNm when biased. The motor has 200 steps/revolution and an in-built resolver with positional accuracy of 60~arcmins (better than 1 step). The in-built temperature sensor in the motor coil is monitored and down-linked via telemetry by the spacecraft thermal control system. 

The motor has a redundant coil to ensure reliability, and the bearing selection has been done to meet the stringent lifetime requirements. Low out-gassing inorganic lubricants and adhesives have been used to construct the motor to minimize molecular contamination. Furthermore, the motors are pre-baked to 200 \textsuperscript{$\circ$}C before assembly into the instrument to reduce further out-gassing during launch and transit. 

The shutter mechanism is directly controlled by the Timing Electronics (TE) board in the SUIT Electronics package, which also controls the CCD clocking (See Section \ref{S-PEPackage}). The TE provides the exposure duration and start/stop pulse signal to the shutter controller circuit that generates the micro-stepping drive profile and scales it for the required exposure duration. The motor is operated in open-loop control, and the resolver feedback is logged into telemetry. 

\subsection{Focusing Mechanism} \label{S-Focusing_Mechanism}

The focusing mechanism shown in the left panel of Figure~\ref{fig:lens} consists of a lens assembly and a linear stage that is driven by an LT20D non-magnetic piezo-electric motor from PiezoMotor Uppsala AB, Sweden. The piezo-motor drives the linear stage on which the lens assembly is mounted. The stage can be moved along the optical axis within  $\pm$~3~mm in steps of $\approx 5.7 \mu m$. The linear stage has a linear encoder that gives relative position accuracy of $\approx 0.5 \mu m$.

The lens assembly is fixed $\approx 228 \mu m$ away from the best focus position during the alignment of \suit on-ground. A Shape Memory Alloy (SMA) type pin-puller mechanism from TiNi Aero-space keeps the linear stage latched in position during launch. The pin-puller is deployed after launch, during the in-flight calibration period, by ground command via the S/C OBC. The decision to move the focusing mechanisms is taken based on the images recorded during the initial calibration phase. During periodic calibrations of the instrument, the focusing mechanism may compensate for the focus shift caused by drift in the thermal control of the instrument.

The drive signal and profile of the piezo-motor are generated by a controller on the mechanisms electronics PCB in the PE box. The command to operate the movement (direction and distance) of the piezo motor is given to the controller by FPGA on the PE box. The PE FPGA either passes the ground commands or uses on board calibration mode commands to operate the piezo motor. The controller for the piezo-motor has been designed and fabricated by the Control Systems Group at URSC.

\section{SUIT Electronics Package}\label{S-PEPackage}

\begin{figure}[!ht]
\centering
\includegraphics[width=1.0\textwidth]{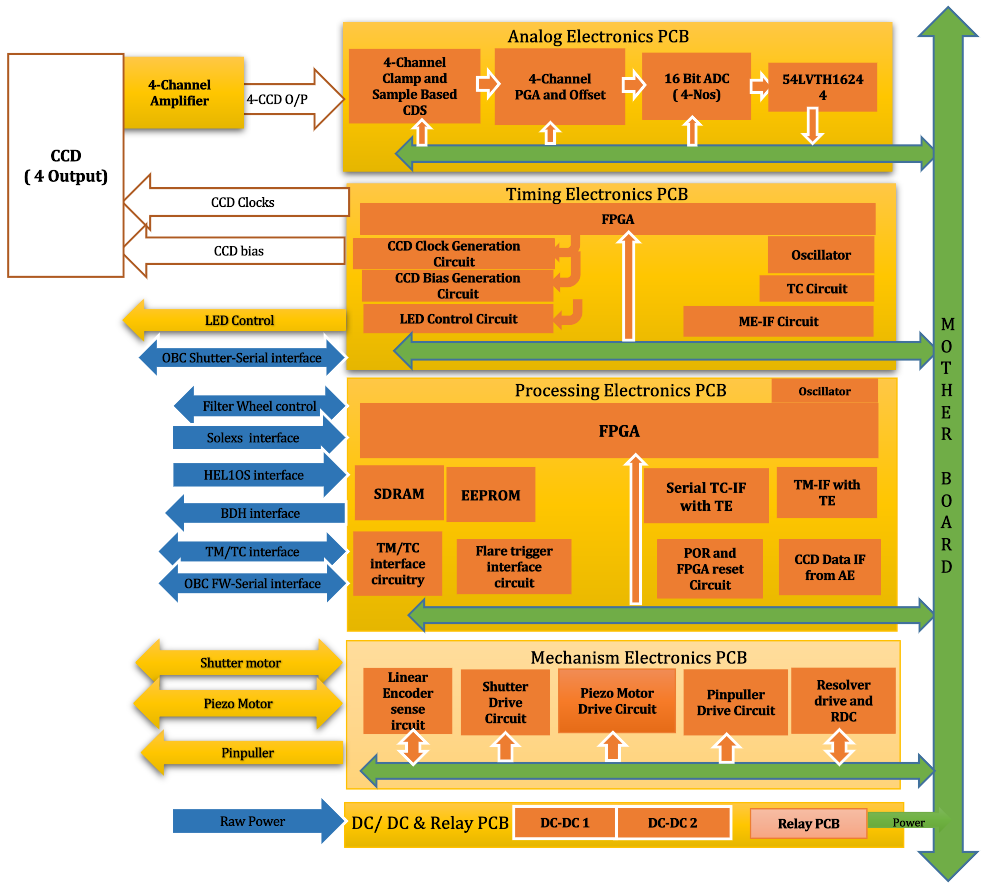}
\caption{Block diagram of the SUIT electronics showing the six PCBs and their interfaces with various internal and external subsystems.}\label{fig:SUIT_EP} 
\end{figure}

The SUIT electronics package consists of six printed circuit boards (PCBs), viz. the Analog Electronics (AE), the Timing Electronics (TE), the Processing Electronics (PE), the Mechanism Electronics (ME), the DC-DC converter and relay board, and the motherboard. Figure~\ref{fig:SUIT_EP} shows the block diagram of the electronics package with various interfaces. 

\subsection{Processing Electronics (PE) Board}\label{S-PEBoard}

\begin{figure}[h!]
\centering
\includegraphics[angle=0, width=1\textwidth]{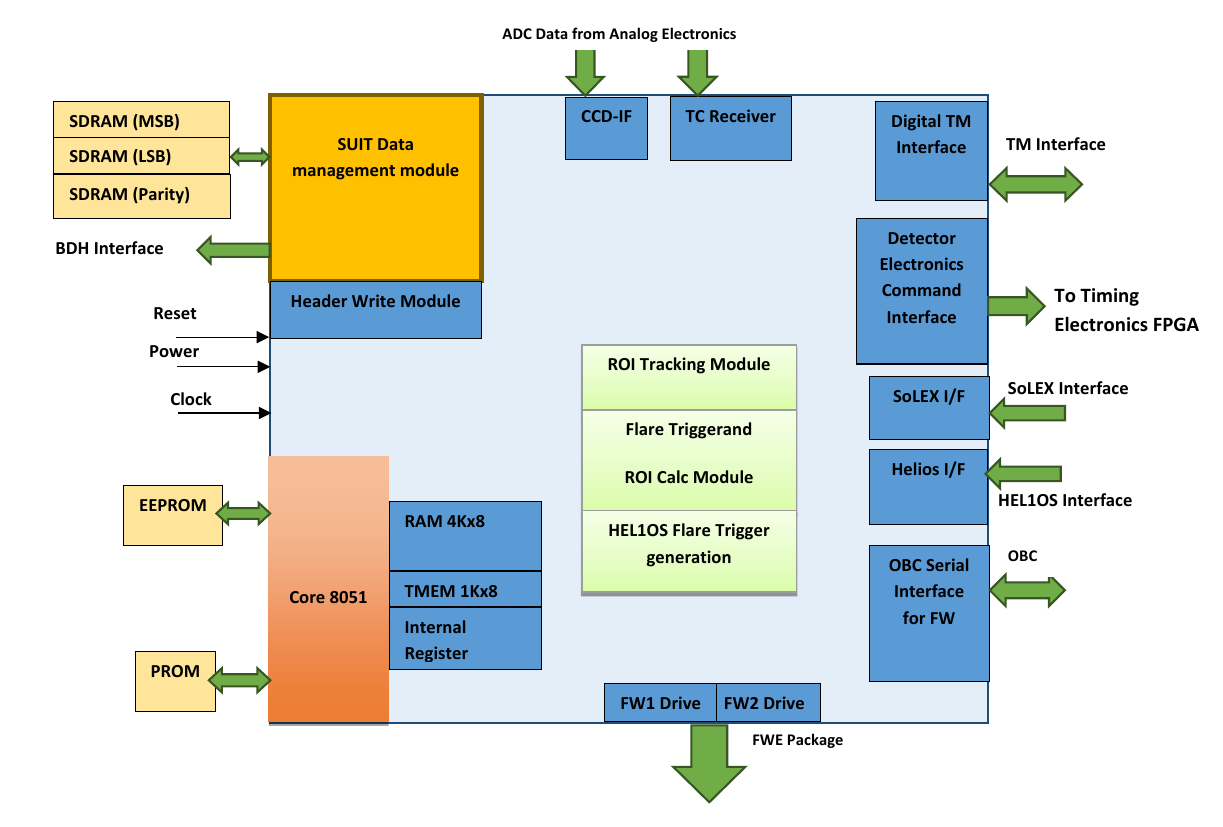}
\caption{Top-level Block Schematic of the SUIT Processing Electronics board.}
\label{fig:PEBoard}       
\end{figure}

The Processing Electronics (PE) board is built around Microsemi make space-qualified FPGA, the main central processor and Synchronous Dynamic Random Access Memory (SDRAM) as the data buffer. Figure~\ref{fig:PEBoard} shows the block schematic of the PE board with its interfaces with other PCBs, telescope and FWE Packages, other instruments, and S/C sub-systems. 

The PE firmware is a combination of native VHDL module(s) and on-chip 8051 IP core microcode. 8051 IP core acts as a master controller that controls and monitors all the operations. SUIT operational sequence for each observation mode is an executable program (called program sequence) built using a set of custom instructions. A set of such program sequences, along with all the configurable parameters, are stored in an E$^2$PROM, and will be copied into FPGA internal registers/memory during operation. All internal and external memory is mapped to the external memory space of Core 8051. This enables the user to upload either a new program sequence(s) or configurable parameter(s) using a simple and compact ground command. At any given time, the ground command loads one program sequence from on-board E$2$PROM into the FPGA internal memory and then triggers its execution.  Time-tagged telecommands are used to select and execute the desired program sequence at a pre-decided time and date.
During the execution of the program sequence, based on the type of instruction, either the master controller executes it or triggers the associated VHDL module(s). For time-bound and event-based operations, some modules automatically run based on the internal state of the state machine/state of internal flag(s) set by the VHDL module, external interface, or telecommand. The firmware design also ensures a high degree of autonomy, allowing nominal instrument operations autonomously based on pre-programmed modes. To execute instrument operations, the PE interacts with AE, TE, ME boards and the FWDE.

The PE is also programmed to autonomously respond to on-board triggers generated either internally or received from other instruments \citep[see][]{Varma_flare}. This on-board detection of a flare event can be ascertained by one/ more of these three independent methods: external flare trigger received from {\sol}, external trigger generated by processing {\hel} data, and internal trigger generated periodically by specific program sequence instructions. Based on the flare trigger, in-built algorithm \citep[see][]{Varma_flare} computes the flare location on Sun's disk to define the RoI, which is used for observation in the Flare mode. The tracking module runs in the background and corrects for shifts in the RoI at a fixed interval. In the flare mode, ADU counts are monitored on the fly, and exposure time is automatically adjusted to maximize the SNR and prevent the CCD from saturation.  

Finally, the PE has serial interfaces to receive tele-commands from the ground station, to send back the housekeeping (telemetry) information at fixed intervals, and to send acquired images along with the metadata.

\subsection{Analogue Electronics (AE), Timing Electronics (TE) and Mechanism Electronics (ME) Boards} \label{S-AE&TEBoards}

The AE and TE boards are primarily responsible for detector operations (i.e., biases, clocks, flushing, readout, etc). The exposure duration and signal to start the exposure are given to AE by the PE board. The AE board generates the clocks and biases required for the CCD readout and relays the digitized data to the PE. The TE board provides the trigger for the shutter mechanism to the ME board. This ensures that the shutter operation is synchronized with CCD timing for accurate exposure control. The TE board also has the drive circuit for the LEDs.  The control signal for the focusing mechanism comes from the PE, while the signals for shutter operations form the TE board. The ME board has two independent circuits for the shutter operations and the focusing mechanism. 

\subsection{Motherboard, DC-DC and Relay Board}\label{S-Motherboard&Power}
The Motherboard has the data and control signal buses for the internal communication between the various boards. It also has the buses for the power supply to each board from the DC-DC and relay board. The DC-DC converters convert the raw bus power received from the S/C, and a relay board has power electronics for signal conditioning and power recycling. The total power budget (raw bus) of the instrument is 35~W.

\section{Structural Design and Analysis}\label{S-Structural}
\begin{figure}[!ht]
\centering
\includegraphics[angle=270,width=0.5\textwidth]{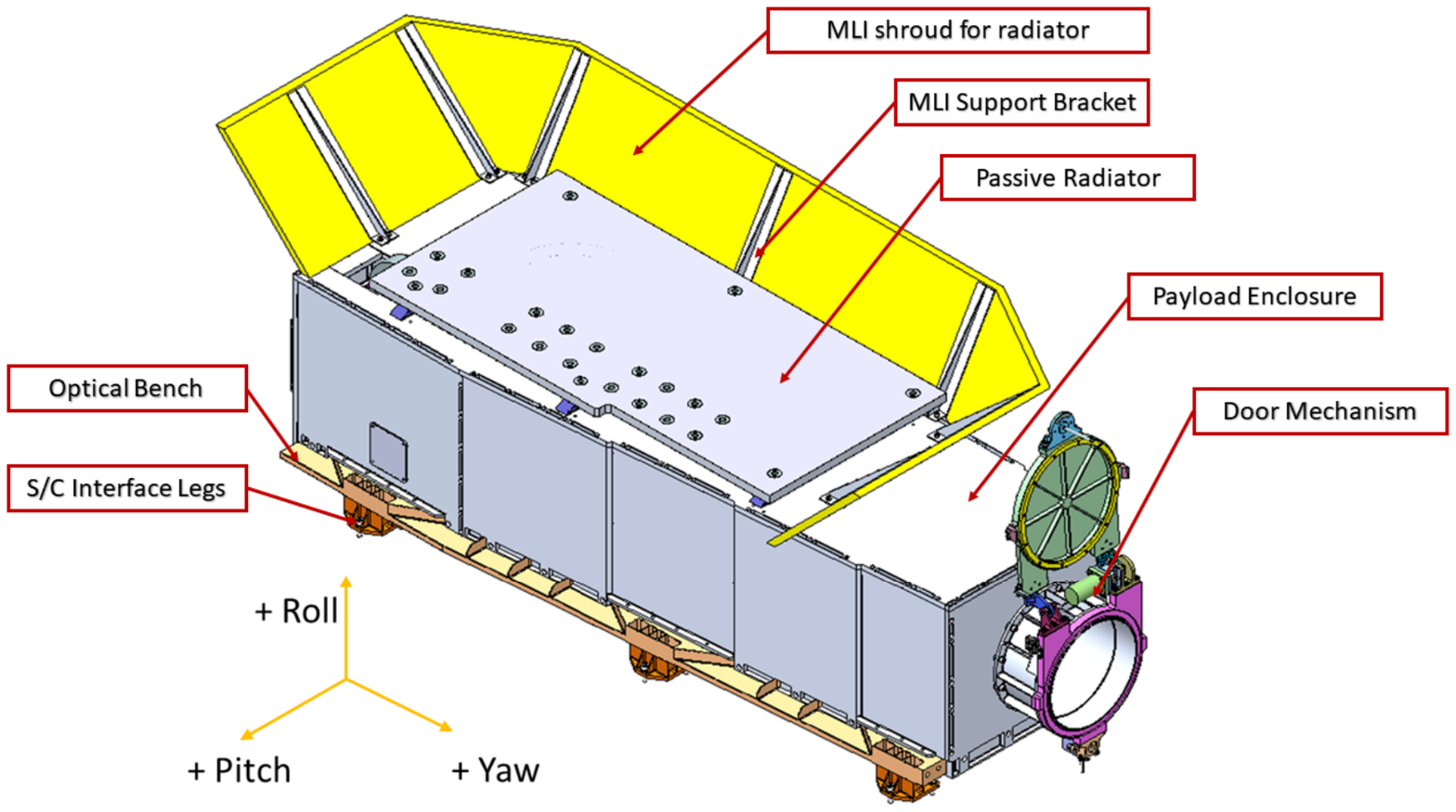}
\includegraphics[angle=270,width=0.45\textwidth]{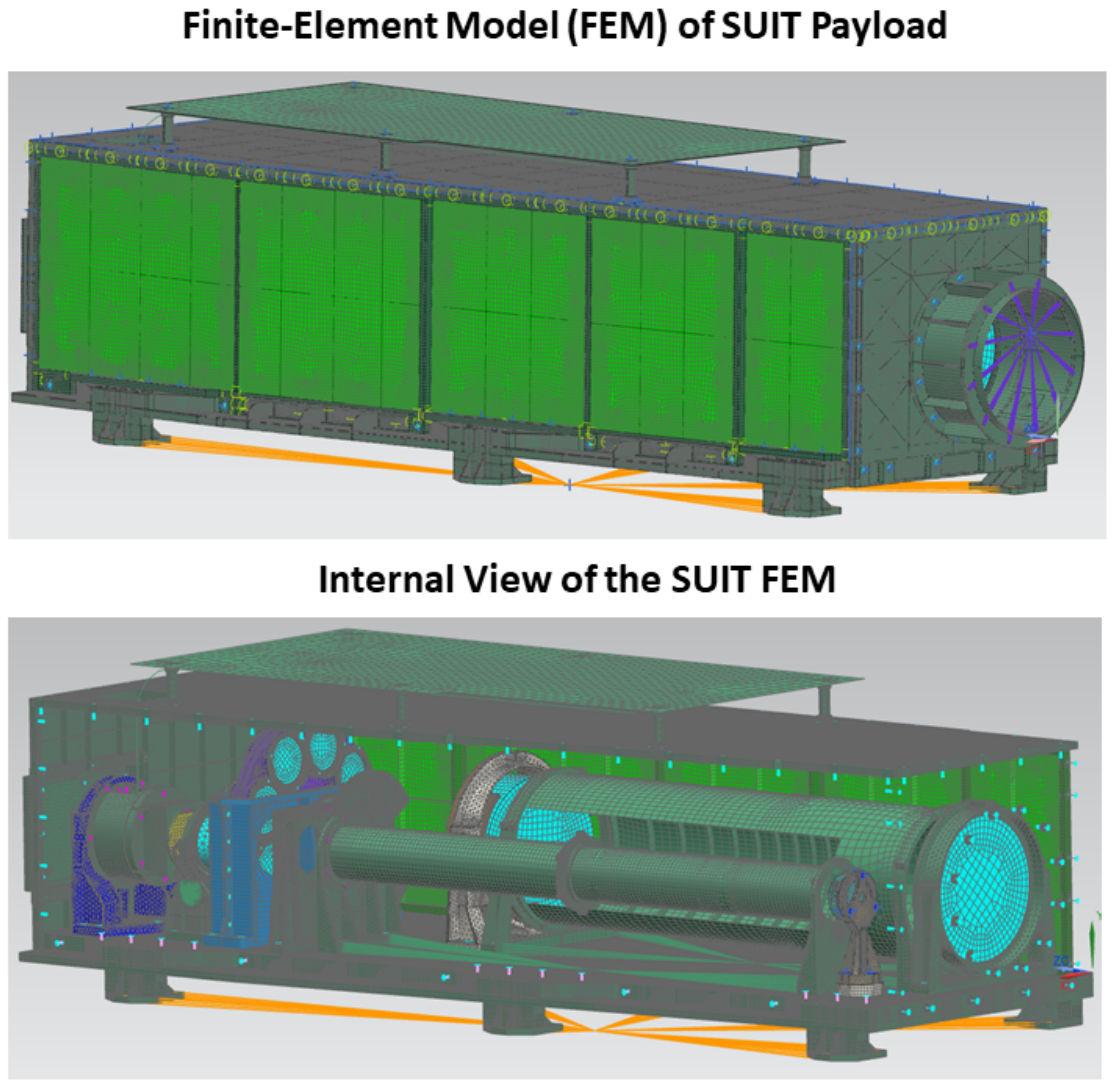}
\caption{Top-level Isometric CAD view of the \suit package (left) and the finite element model of \suit showing the mesh elements for the full model (right top) and the internal view with FEM meshes for the assemblies (right bottom).}
\label{fig:ISOView}       
\end{figure}

Figure~\ref{fig:ISOView} shows the isometric CAD view of the telescope assembly. The instrument optical bench, a lightweight plate made of Titanium alloy (Ti-6Al-4V), serves as the structural support element for all sub-assemblies of the telescope. It also acts as a metering structure for the opto-mechanical system. It maintains the relative positions of the optical elements within the desired tolerance range due to the low thermal expansion coefficient. The optical bench is interfaced with the S/C top deck through six mounting legs made of Titanium alloy. 

The enclosure for \suit is made of lightweight Aluminium alloy (Al 6061-T6) panels that provide a light-leak-proof optical cavity. The enclosure also supports the thermal control elements, including the insulation, patch heaters, and passive radiator. Brackets mounted on the enclosure support the MLI insulation shroud that shields the passive radiator from direct illumination from the Sun and thermal load from S/C surfaces.   

The structural design and analysis of the instrument were carried out at IUCAA in collaboration with the structures team at the URSC. A detailed Finite Element (FE) model was used in NX Nastran to carry out the structural analysis at the instrument level. The detailed FE Model, shown in Figure~\ref{fig:ISOView}, has more than 900,000 nodes to accurately simulate the modal behavior, deformation, and stresses due to launch loads. A coarse FE model of the instrument with 160,000 nodes was created and was integrated into the spacecraft (S/C) FEM model for S/C level analysis of launch loads and modal behavior. The results from three different levels of structural analysis were used to optimize the structure within the available mass budget. The mechanical and opto-mechanical design of individual sub-assemblies were optimized based on the instrument level structural analysis results.

\section{Thermal Control System}\label{S-Thermal}
\begin{table*}[!ht]
\caption{Thermal design parameters for \suit.} \label{tab:Therm_req}
\begin{tabular}{lccc}
\hline
\textbf{Elements}		&	\textbf{Operational} 			&	\textbf{Non-Operational}&	\textbf{Temporal}\\
 &	\textbf{Temp (\textsuperscript{$\circ$}C)}	& \textbf{Temp (\textsuperscript{$\circ$}C)} & \textbf{Stability (\textsuperscript{$\circ$}C)} \\
\hline
Optical Bench and 	&	20$\pm$3		&	0{--}40			& $\pm$ 1 \\
optical elements		&					&					&			\\
Detector 			& -55$\pm$3		& {--}75{--}80 		& $\pm$ 1\\
Thermal Filter 		& 5{--}50 			& {--}20{--}75		& {--} \\
Science Filters 		& 20$\pm$5 			& 0{--}40 			& {--}\\
Filter Wheel Motor 	& 15{--}40 			& {--}5{--}80 		& {--} \\
Shutter Motor 		& 15{--}40 			& {--}5{--}80 		& {--}\\
Entrance Baffle 		& {--}10{--}50 		& {--}35{--}75 		& {--}\\
Payload Electronics	& {--}5{--}40 		& {--}30{--}65 		& {--}\\
\hline
\end{tabular}	
\end{table*}
Temperature sensors have been mounted on the cover panels, optical bench and critical components like motors and the detector assembly to estimate the actual heat load at different locations across the optical cavity. Regulation of the heat load has also been dealt with by using patch heaters or radiator plates. In this perspective, it is imperative to mention that the cover panels, entrance door cover, external baffle and the entire S/C deck have multi-layer insulation (MLI) blankets on the outside to insulate the instrument from the space environment (at {-}269~\textsuperscript{$\circ$}C) and from the solar radiation energy ($\approx$~1439 W/m\textsuperscript{2}). Furthermore, the optical bench is isolated from the S/C deck by 30~mm, using legs with high thermal resistance and rigidity. 

The coldest part along the optical bench is the CCD, which is to be maintained at -55~\textsuperscript{$\circ$}C, to ensure the dark current at $\approx$~4~e~pix$^{-1}$s$^{-1}$. This is achieved by mounting a passive radiator, a honeycomb panel with a total area of 2500~cm\textsuperscript{2}, on the top cover panel with insulating legs. An MLI shroud is used to shade the radiator for the direct and reflected (from walls of the neighboring instrument) sunlight. This radiator serves as a heat sink for the CCD via a flexible gold-coated copper cold finger connected with an ethane heat pipe.

\section{Data Analysis Pipeline, Management and Distribution}\label{S-Data_Management}
The SUIT operational modes are designed to limit the daily data generated to within the downlink budget/ downlink limit, which is ~20 GBytes per day. The data limit, along with exposure duration, CCD readout duration, filter wheel movement and dwell time, filter sequencing, etc., determines the maximum cadence for the filters in a particular mode. The SUIT PE converts the CCD readout into an image frame for each exposure and attaches header information, including exposure duration, filter ID, timestamp, mode, etc. In the synoptic normal mode that consists of recording full disk and partial disk images of the Sun, SUIT will generate approximately ~40 GBytes of science data every day. The data is relayed to the S/C BDH subsystem, which applies a factor of two lossless compression and stores the data in a solid-state memory.

The science data will be downloaded daily, along with the other instruments on board {\aditya} and S/C auxiliary data, as a binary stream via S-band link to the ISRO Telemetry, Tracking and Command Network (ISTRAC) Ground station and relayed to the ISRO Space Science Data Center (ISSDC) for further processing. The compressed binary data (raw Data) is decompressed and segregated into Level-0 data packets for each instrument. Each level-0 binary format data packet for SUIT will have data for 30 minutes of observations, including separated images in binary format and auxiliary data (AD).

The processing of level-0 binary files is done in a pipeline software developed by the instrument team and hosted at the payload operations centre (POC) at IUCAA, which converts the binary image files into FITS format, extracts relevant parameters from S/C telemetry and adds them to the FITS header. A universal time (UT) is also added to the FITS header using the on board timestamp for each image that is included in the image header and the time correlation information in the Time Correlation Information (TCI) file.

The level-1 processing at POC applies basic instrument response corrections viz. dark, flat, scatter, hot pixels corrections, etc, to the images. Additionally, the Orbit, Attitude and Time (OAT) files are used to identify the location of the solar north pole in each image. The orientation of the disk is corrected, and the information for heliocentric coordinate mapping is included in the header file. Level 1 data is fully science-ready data. The Level-2 processing produces data products that will be fully calibrated for radiometric calibration. The POC will send these data products to ISSDC for archival and dissemination through a dedicated web portal for the scientific investigators and the community for further analysis and outreach. 

\section{Conclusions}\label{S-conclusions}
\suit is designed to provide full disk and region-of-interest images of the Sun in near and mid ultraviolet wavelength range \textit{viz.} 200{--}400~nm. The observations recorded by SUIT will help us understand the dynamic coupling of the solar atmosphere and the mass and energy transfer within. Combined with other space and ground-based observatories, SUIT observations will provide seamless coverage of the solar atmosphere. Moreover, for the first time, it will allow us to measure and monitor the spatially resolved solar features contributing to changes in the solar spectral irradiance in the near and mid ultraviolet, which is a critical missing ingredient in modeling the influence of the Sun on climate. 

\begin{acks}
We thank the referee for constructive comments and suggestions. SUIT is built by a consortium led by the Inter-University Centre for Astronomy and Astrophysics (IUCAA), Pune, and supported by ISRO as part of the Aditya-L1 mission. The consortium comprises CESSI-IISER Kolkata (MoE), IIA, MAHE, MPS, USO/PRL, and Tezpur University. Aditya-L1 mission was conceived and realised with the help from all ISRO Centres and instruments were realised by the instrument PI Institutes in close collaboration with ISRO and many other national institutes - Indian Institute of Astrophysics (IIA); Inter University Centre of Astronomy and Astrophysics (IUCAA); Laboratory for Electro-optics System (LEOS) of ISRO; Physical Research Laboratory (PRL); U R Rao Satellite Centre of ISRO; Vikram Sarabhai Space Centre (VSSC) of ISRO. SP acknowledges the Manipal Centre for Natural Sciences, Centre of Excellence, and Manipal Academy of Higher Education (MAHE) for facilities and support. 
\end{acks}
\begin{authorcontribution}
All authors have contributed significantly.
\end{authorcontribution}
\begin{fundinginformation}
Aditya-L1 is an observatory class mission which is funded and operated by the Indian Space Research Organization. DT acknowledges support from the Max-Planck India partner groups on Coupling and Dynamics of the solar atmosphere at IUCAA. SKS acknowledges funding from the European Research Council (ERC) under the European Union's Horizon 2020 research and innovation programme (grant agreement No. 101097844 — project WINSUN).
\end{fundinginformation}

%
%
%

\bibliographystyle{spr-mp-sola}
\bibliography{sola_bibliography_example}  
\end{document}